\documentclass{article}

\usepackage[preprint]{neurips_2020}
\usepackage[utf8]{inputenc} % allow utf-8 input
\usepackage[T1]{fontenc}    % use 8-bit T1 fonts
\usepackage{hyperref}       % hyperlinks
\usepackage{url}            % simple URL typesetting
\usepackage{booktabs}       % professional-quality tables
\usepackage{amsfonts}       % blackboard math symbols
\usepackage{nicefrac}       % compact symbols for 1/2, etc.
\usepackage{microtype}      % microtypography
\usepackage{amsmath, amssymb}
\usepackage{amsthm}

\usepackage{subfigure}
\usepackage{graphicx}
\usepackage{float}

\usepackage{microtype}
\usepackage{booktabs} % for professional tables
\usepackage{multirow} % added by myself
\usepackage{makecell} % added by myself
\usepackage{algorithm,algorithmicx,algpseudocode}

\usepackage{amsmath}
\usepackage{amssymb}
\usepackage{mathtools}
\usepackage{amsthm}

\usepackage[capitalize,noabbrev]{cleveref}
\usepackage[OT2, T1]{fontenc}
\theoremstyle{plain}

\theoremstyle{definition}

\theoremstyle{remark}

\usepackage[textsize=tiny]{todonotes}

\allowdisplaybreaks[1] % this is added by us.

\title{Measurement-conditioned Denoising Diffusion Probabilistic Model for Under-sampled Medical Image Reconstruction}

\begin{document}

\author{%
  Yutong Xie\\
  Academy for Advanced Interdisciplinary Studies\\
  Peking University\\
  Beijing, 100871 \\
  % examples of more authors
  \And
  Quanzheng Li \\
  MGH/BWH Center for Advanced Medical Computing and Analysis\\ Gordon Center for Medical Imaging, Department of Radiology\\
  Massachusetts General Hospital and Harvard Medical School\\
  Boston, MA 02114\\
  \texttt{li.quanzheng@mgh.harvard.edu} \\
}

\maketitle

\begin{abstract}
We propose a novel and unified method, measurement-conditioned denoising diffusion probabilistic model (MC-DDPM), for under-sampled medical image reconstruction based on DDPM. Different from previous works, MC-DDPM is defined in measurement domain (e.g. k-space in MRI reconstruction) and conditioned on under-sampling mask. We apply this method to accelerate MRI reconstruction and the experimental results show excellent performance,  outperforming full supervision baseline and the state-of-the-art score-based reconstruction method. Due to its generative nature, MC-DDPM can also quantify the uncertainty of reconstruction. Our code is available on github\footnote{https://github.com/Theodore-PKU/MC-DDPM}.
\end{abstract}

\section{Introduction}
\label{sec:introduction}

Reconstruction from under-sampled measurements in medical imaging has been deeply studied over the years, including reconstruction of accelerated magnetic resonance imaging (MRI) \citep{aggarwal2018modl,hammernik2018learning,eo2018kiki,han2019k}, sparse view or limited angles computed tomography (CT) \citep{han2018framing,zhang2019jsr,wang2019admm} and digital breast tomosynthesis (DBT). Most of works aim to obtain one sample of the posterior distribution $p\left(\mathbf{x} \mid \mathbf{y} \right)$ where $\mathbf{x}$ is the reconstructed target image and $\mathbf{y}$ is the under-sampled measurements.

Recently, denoising diffusion probabilistic models (DDPM) \citep{sohl2015deep,ho2020denoising}, which is a new class of unconditional generative model, have demonstrated superior performance and have been widely used in various image processing tasks. DDPM utilizes a latent variable model to reverse a diffusion process, where the data distribution is perturbed to the noise distribution by gradually adding Gaussian noise. Similar to DDPM, score-based generative models \citep{hyvarinen2005estimation,song2019generative} also generate data samples by reversing a diffusion process. Both DDPM and score-based models are proved to be discretizations of different continuous stochastic differential equations by \citep{song2020score}. The difference between them lies in the specific setting of diffusion process and sampling algorithms. They have been applied to the generation of image \citep{song2020score,nichol2021improved,dhariwal2021diffusion}, audio \citep{kong2020diffwave} or graph \citep{niu2020permutation}, and to conditional generation tasks such as in in-painting \citep{song2019generative,song2020score}, super-resolution \citep{choi2021ilvr,saharia2021image} and image editing \citep{meng2021sdedit}. In these applications, the diffusion process of DDPM or score-based generative model is defined in data domain, and is unconditioned although the reverse process could be conditioned given certain downstream task. Particularly, the score-based generative model has been used for under-sampled medical image reconstruction \citep{jalal2021robust,song2021solving,chung2021score}, where the diffusion process is defined in the domain of image $\mathbf{x}$ and is irrelevant to under-sampled measurements $\mathbf{y}$. 
 
\begin{figure}[t]
    \centering
    \includegraphics[width=\linewidth]{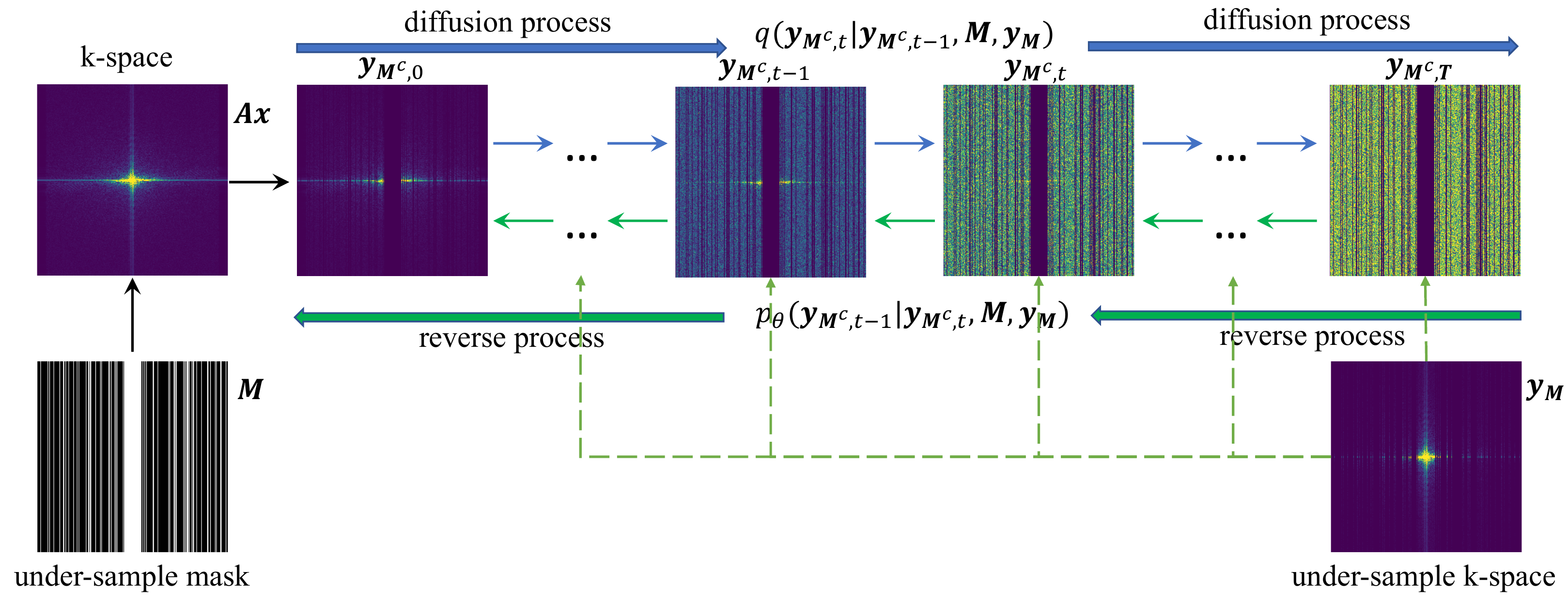}
    \caption{Overview of the proposed method illustrated by the example of under-sampled MRI reconstruction. Diffusion process: starting from the non-sampled k-space $\mathbf{y}_{\mathbf{M}^c, 0}$, Gaussian noise is gradually added until time $T$. Reverse process: starting from total noise, $\mathbf{y}_{\mathbf{M}^c, 0}$ is generated step by step. The details of notations is presented in Sect.~\ref{sec:method}.}
    \label{fig:method}
\end{figure}

In this paper, We design our method based on DDPM rather than score-based generative model because DDPM is more flexible to control the  noise distribution. We propose a novel and unified method, measurement-conditioned DDPM (MC-DDPM) for under-sampled medical image reconstruction based on DDPM (Fig.\ref{fig:method} illustrates the method by the example of under-sampled MRI reconstruction), where the under-sampling is in the measurement space (e.g. k-space in MRI reconstruction) and thus the conditional diffusion process is also defined in the measurement sapce.  Two points distinguish our method from previous works \citep{jalal2021robust,song2021solving,chung2021score}: (1) the diffusion and sampling process are defined in measurement domain rather than image domain; (2) the diffusion process is conditioned on under-sampling mask so that data consistency is contained in the model naturally and inherently, and there is no need to execute extra data consistency when sampling. The proposed method allows us to sample multiple reconstruction results from the same measurements $\mathbf{y}$. Thus, we are able to quantify uncertainty for $q\left(\mathbf{x}\mid \mathbf{y}\right)$, such as pixel-variance. Our experiments on accelerated MRI reconstruction show MC-DDPM can generate samples of high quality of $q\left(\mathbf{x}\mid \mathbf{y}\right)$ and it outperforms baseline models and proposed method by \citep{chung2021score} in evaluation metrics. 
% In addition, utilizing the method proposed in \citep{xie2022trained} we also trained a network to directly estimate pixel-variance and compare its results to the one estimated by MC-DDPM. The closeness of two estimations further verify our reconstruction method.

This paper is organized as follows: relevant background on DDPM and the under-sampled medical image reconstruction task is in Sect.~\ref{sec:background}; details of the proposed method MC-DDPM is presented in Sect~\ref{sec:method}; specifications about the implementation of the application to accelerated MRI reconstruction, experimental results and discussion are given in Sect.~\ref{sec:experiments}; 
% discussion about MCDDMPM is presented in Sect.~\ref{sec:discussion}; 
% we show the related works in Sect.~\ref{sec:related_works} and conclude our work in Sect.~\ref{sec:conclusion}. 
we conclude our work in Sect.~\ref{sec:conclusion}.

\section{Background}
\label{sec:background}

\subsection{Denoising Diffusion Probabilistic Model}
\label{sec:background_ddpm}

DDPM \citep{ho2020denoising} is a certain parameterization of diffusion models \citep{sohl2015deep}, which is a class of latent variable models using a Markov chain to convert the noise distribution to the data distribution. It has the form of $p_{\theta}\left(\mathbf{x}_{0}\right):= \int p_{\theta} \left( \mathbf{x}_{0:T} \right) \mathrm{d} \mathbf{x}_{1:T} $, where $\mathbf{x}_{0}$ follows the data distribution $q\left(\mathbf{x}_{0} \right)$ and $\mathbf{x}_{1}, ..., \mathbf{x}_{T}$ are latent variables of the same dimensionality as $\mathbf{x}_{0}$. The joint distribution $p_{\theta}\left( \mathbf{x}_{0:T} \right)$ is defined as a Markov chain with learned Gaussian transitions starting from $p\left(\mathbf{x}_{T} \right)=\mathcal{N}\left(\mathbf{0}, \mathbf{I} \right)$:
\begin{equation}
\label{eq:ddpm_p}
    p_{\theta}\left(\mathbf{x}_{0: T}\right) := p\left(\mathbf{x}_{T}\right) \prod_{t=1}^{T} p_{\theta}\left(\mathbf{x}_{t-1} \mid \mathbf{x}_{t}\right), p_{\theta}\left(\mathbf{x}_{t-1} \mid \mathbf{x}_{t}\right):=\mathcal{N} \left(\boldsymbol{\mu}_{\theta}\left(\mathbf{x}_{t}, t\right), \sigma_{t}^{2} \mathbf{I}\right).
\end{equation} 
The sampling process of $p_{\theta}\left( \mathbf{x}_{0}\right)$ is: to sample $\mathbf{x}_{T}$ from $\mathcal{N}\left(\mathbf{0}, \mathbf{I} \right)$ firstly; then,  to sample $\mathbf{x}_{t-1}$ from $p_{\theta}\left(\mathbf{x}_{t-1} \mid \mathbf{x}_{t}\right)$ until $\mathbf{x}_{0}$ is obtained. It can be regarded as a reverse process of the diffusion process, which converts the data distribution to the noise distribution $\mathcal{N} \left(\mathbf{0}, \mathbf{I} \right)$. In DDPM the diffusion process is fixed to a Markov chain that gradually adds Gaussian noise to the data according to a variance schedule $\beta_1$, $...$, $\beta_T$:
\begin{equation}
\label{eq:ddpm_diffusion_process}
    q\left(\mathbf{x}_{1:T} \mid \mathbf{x}_{0}\right) := \prod_{t=1}^T q\left( \mathbf{x}_{t} \mid \mathbf{x}_{t-1}\right), \quad q\left(\mathbf{x}_{t} \mid \mathbf{x}_{t-1} \right):= \mathcal{N}\left( \alpha_{t} \mathbf{x}_{t-1}, \beta_{t}^2 \mathbf{I} \right),
\end{equation}
where $\alpha_t^2 + \beta_t^2 = 1$ for all $t$ and $\beta_1$, $...$, $\beta_T$ are fixed to constants and their value are set specially so that $q\left( \mathbf{x}_{T} \mid \mathbf{x}_{0} \right) \approx \mathcal{N} \left(\mathbf{0}, \mathbf{I} \right)$. 
% The training of $p_{\theta}\left(\mathbf{x}_{0:T}\right)$, i.e. $\boldsymbol{\mu}_{\theta}\left(\cdot, t\right)$, is performed by optimizing the variational bound on negative log likelihood. By neglecting constants and reweighting, the training objective is transformed to a denoising score-matching loss. The details of derivation of training objective and the sampling algorithm can be seen in supplementary materials. 

\subsection{Under-sampled Medical Image Reconstruction}
\label{sec:background_med_recon}

Suppose $\mathbf{x} \in \mathbb{R}^n$ represents a medical image and $\mathbf{y} \in \mathbb{R}^m, m < n$ is the under-sampled measurements which is obtained by the following forward model:
\begin{equation}
\label{eq:undersampled_forward}
    \mathbf{y} =\mathbf{P}_{\Omega} \mathbf{A} \mathbf{x} + \boldsymbol{\epsilon}, 
\end{equation}
where $\mathbf{A} \in \mathbb{R}^{n \times n}$ is the measuring matrix and usually is invertible, $\mathbf{P}_{\Omega} \in \mathbb{R}^{m\times n}$ is the under-sampling matrix with the given sampling pattern $\Omega$,\footnote{Assuming the sampling pattern $\Omega$ is $\{s_1, ..., s_m\} \subseteqq \{1, ..., n\}$, the element of $\mathbf{P}_{\Omega}$ at position $(i, s_i)$, $i = 1, ..., m$, is $1$ and other elements are all $0$.} and $\boldsymbol{\epsilon}$ is the noise. For example, $\mathbf{x}$ is a CT image, $\mathbf{A}$ is the Radon transform matrix and $\mathbf{y}$ is the sinogram of limited angles. Under-sampled medical image reconstruction is to reconstruct $\mathbf{x}$ from $\mathbf{y}$ as possible. Assuming $\mathbf{x}$ follows a distribution of $q\left( \mathbf{x} \right)$ and given $\mathbf{P}_{\Omega}$, according to Bayesian Formula, we can derive the posterior distribution as follows (usually $\mathbf{P}_{\Omega}$ is neglected):
\begin{equation}
\label{eq:bayesian_formula}
    q \left( \mathbf{x} \mid \mathbf{y}, \mathbf{P}_{\Omega} \right) = \frac{q \left(\mathbf{x}, \mathbf{y} \mid  \mathbf{P}_{\Omega} \right)}{q\left( \mathbf{y} \right)} = \frac{q\left(\mathbf{y} \mid \mathbf{x}, \mathbf{P}_{\Omega} \right) q \left(\mathbf{x} \right)}{q\left( \mathbf{y} \right)}.
\end{equation}
Therefore, the task of under-sampled medical image reconstruction to reconstruct the posterior distribution.

\section{Method: Measurement-conditioned DDPM}
\label{sec:method}

Inspired by DDPM, we propose measurement-conditioned DDPM (MC-DDPM) which is designed for under-sampled medical image reconstruction. In this section, we formulate the MC-DDPM, including the diffusion process and its reverse process, training objective and sampling algorithm. For the convenience, we use new notations different from Eq.~\ref{eq:undersampled_forward} to represent the under-sampled forward model:
\begin{equation}
\label{eq:undersampled_forward_new}
    \mathbf{y}_{\mathbf{M}} = \mathbf{M} \mathbf{A} \mathbf{x} + \boldsymbol{\epsilon}_{\mathbf{M}},
\end{equation}
where $\mathbf{M} \in \mathbb{R}^{n \times n}$ is a  diagonal matrix whose diagonal elements are either $1$ or $0$ depending on the sampling pattern $\Omega$.\footnote{Specifically, $M_{i, i} = 1$ if $i \in \Omega$. Otherwise its value is $0$.} $\mathbf{y}_{\mathbf{M}}$ and $\boldsymbol{\epsilon}_{\mathbf{M}}$ are both $n$-dimension vectors and their components at non-sampled positions are $0$. The merit of the new notations is that we can further define $\mathbf{M}^{c} = \mathbf{I} - \mathbf{M}$ (the superscript $c$ means complement) and $\mathbf{y}_{\mathbf{M}^c} = \mathbf{M}^{c} \mathbf{A} \mathbf{x}$ which represents the non-sampled measurements. In this paper, we assume $\boldsymbol{\epsilon}_{\mathbf{M}} = \mathbf{0}$. Then, we have $\mathbf{y}_{\mathbf{M}} + \mathbf{y}_{\mathbf{M}^c} = \mathbf{A} \mathbf{x} $, i.e. $\mathbf{y}_{\mathbf{M}} + \mathbf{y}_{\mathbf{M}^c}$ is the full-sampled measurements.  In addition, the posterior distribution of reconstruction can be rewritten as $q \left(\mathbf{x} \mid \mathbf{M}, \mathbf{y}_{\mathbf{M}} \right)$. Through this paper, the subscript $\mathbf{M}$ or $\mathbf{M}^c$ in notations indicates that only components at under-sampled or non-sampled positions  are not $0$.

% \subsection{Formation}
% \label{sec:method_formation}

The purpose of reconstruction task is to estimate $q \left( \mathbf{x} \mid \mathbf{M}, \mathbf{y}_{\mathbf{M}}\right)$.  Since $\mathbf{y}_{\mathbf{M}}$ is known and $\mathbf{x} = \mathbf{A}^{-1} \left( \mathbf{y}_{\mathbf{M}} + \mathbf{y}_{\mathbf{M}^c}\right)$ , the problem is transformed to estimate $q \left(\mathbf{y}_{\mathbf{M}^c} \mid \mathbf{M}, \mathbf{y}_{\mathbf{M}}\right)$. Because $\mathbf{M}$ and $\mathbf{M}^c$ are equivalent as the condition, we can replace  $ q \left(\mathbf{y}_{\mathbf{M}^c} \mid \mathbf{M}, \mathbf{y}_{\mathbf{M}}\right)$ by $ q \left(\mathbf{y}_{\mathbf{M}^c} \mid \mathbf{M}^c, \mathbf{y}_{\mathbf{M}}\right)$. Based on this observation, we propose MC-DDPM which solves the reconstruction problem by generating samples of $ q \left(\mathbf{y}_{\mathbf{M}^c} \mid \mathbf{M}^c, \mathbf{y}_{\mathbf{M}}\right)$. MC-DDPM is defined in measurement domain, instead of image domain as usual DDPM, and is conditioned on the non-sampling matrix $\mathbf{M}^c$ and sampled measurements $\mathbf{y}_{\mathbf{M}}$. It has the following form:
\begin{equation}
\label{eq:MC-DDPM_model}
    p_{\theta}\left(\mathbf{y}_{\mathbf{M}^c, 0} \mid \mathbf{M}^c, \mathbf{y}_{\mathbf{M}} \right):= \int  p_{\theta}\left(\mathbf{y}_{\mathbf{M}^c, 0:T} \mid \mathbf{M}^c, \mathbf{y}_{\mathbf{M}} \right) \mathrm{d} \mathbf{y}_{\mathbf{M}^c, 1:T},
\end{equation}
where $\mathbf{y}_{\mathbf{M}^c, 0}$ = $\mathbf{y}_{\mathbf{M}^c}$. $p_{\theta}\left(\mathbf{y}_{\mathbf{M}^c, 0:T}\mid \mathbf{M}^c, \mathbf{y}_{\mathbf{M}} \right)$ is defined as follows:
\begin{equation*}
\label{eq:MC-DDPM_reverse_process}
\begin{split}
    p_{\theta}\left(\mathbf{y}_{\mathbf{M}^c,0: T}\mid \mathbf{M}^c, \mathbf{y}_{\mathbf{M}}\right) := p\left(\mathbf{y}_{\mathbf{M}^c, T}\mid  \mathbf{M}^c, \mathbf{y}_{\mathbf{M}}\right) \prod_{t=1}^{T} p_{\theta}\left(\mathbf{y}_{\mathbf{M}^c,t-1} \mid \mathbf{y}_{\mathbf{M}^c,t}, \mathbf{M}^c, \mathbf{y}_{\mathbf{M}}\right), \\ p_{\theta}\left(\mathbf{y}_{\mathbf{M}^c, t-1} \mid \mathbf{y}_{\mathbf{M}^c,t}, \mathbf{M}^c, \mathbf{y}_{\mathbf{M}}\right):=\mathcal{N} \left(\boldsymbol{\mu}_{\theta}\left(\mathbf{y}_{\mathbf{M}^c,t}, t, \mathbf{M}^c, \mathbf{y}_{\mathbf{M}}\right), \sigma_{t}^{2} \mathbf{M}^c\right),
\end{split}
\end{equation*}
where $\sigma_{t}^{2} \mathbf{M}^c$ is the covariance matrix and it means the noise is only added at non-sampled positions because for all $t$ the components of $\mathbf{y}_{\mathbf{M}^c,t}$ at under-sampled positions are always $0$. If the conditions $\left( \mathbf{M}^c, \mathbf{y}_{\mathbf{M}} \right)$ in equations above is removed, they degrade to the form of Eq.~\ref{eq:ddpm_p}.

Similar to DDPM, the sampling process of $p_{\theta}\left(\mathbf{y}_{\mathbf{M}^c,0}\mid \mathbf{M}^c, \mathbf{y}_{\mathbf{M}}\right)$ is a reverse process of the diffusion process which is also defined in measurement domain. Specifically, the Gaussian noise is gradually added to the non-sampled measurements $\mathbf{y}_{\mathbf{M}^c, 0}$. The diffusion process has the following form:
\begin{equation}
\label{eq:MC-DDPM_diffusion_process}
\begin{split}
     q\left(\mathbf{y}_{\mathbf{M}^c,1:T} \mid \mathbf{y}_{\mathbf{M}^c,0}, \mathbf{M}^c, \mathbf{y}_{\mathbf{M}}\right) := \prod_{t=1}^T q\left( \mathbf{y}_{\mathbf{M}^c,t} \mid \mathbf{y}_{\mathbf{M}^c,t-1}, \mathbf{M}^c, \mathbf{y}_{\mathbf{M}}\right), \\ q\left(\mathbf{y}_{\mathbf{M}^c,t} \mid \mathbf{y}_{\mathbf{M}^c,t-1}, \mathbf{M}^c, \mathbf{y}_{\mathbf{M}} \right):= \mathcal{N}\left(\alpha_{t} \mathbf{y}_{\mathbf{M}^c,t-1}, \beta_{t}^2 \mathbf{M}^c \right),
\end{split}
\end{equation}
There are two points worthy of noting: (1) $\alpha_t, \beta_{t}$ are not restricted to satisfy $\alpha_t^2 + \beta_t^2 = 1$; (2) formally, we add $\mathbf{y}_{\mathbf{M}}$ as one of the conditions, but it has no effect on the diffusion process in fact. Let $\bar{\alpha}_{t}=\prod_{i=i}^{t} \alpha_{i}, \bar{\beta}_{t}^2={\sum_{i=1}^{t} \frac{\bar{\alpha}_{t}^{2}}{\bar{\alpha}_{i}^{2}} \beta_{i}^{2}}$, and we additionally define $\bar{\alpha}_{0} = 1, \bar{\beta}_{0} = 0$. Then, we can derive that:
\begin{equation}
\label{eq:MC-DDPM_q_t}
    q\left(\mathbf{y}_{\mathbf{M}^c,t} \mid \mathbf{y}_{\mathbf{M}^c,0}, \mathbf{M}^c, \mathbf{y}_{\mathbf{M}} \right) = \mathcal{N}\left(\bar{\alpha}_{t} \mathbf{y}_{\mathbf{M}^c,0}, \bar{\beta}_{t}^{2} \mathbf{M}^c\right),
\end{equation}
\begin{equation}
\label{eq:MC-DDPM_q_t-1_t_0}
    q\left(\mathbf{y}_{\mathbf{M}^c, t-1} \mid \mathbf{y}_{\mathbf{M}^c, t}, \mathbf{y}_{\mathbf{M}^c,0}, \mathbf{M}, \mathbf{y}_{\mathbf{M}}\right) = \mathcal{N}\left(\tilde{\boldsymbol{\mu}}_{t}, \tilde{\beta}_{t}^{2} \mathbf{M}^c\right),
\end{equation}
where $\tilde{\boldsymbol{\mu}}_{t}=\frac{\alpha_{t} \bar{\beta}_{t-1}^{2}}{\bar{\beta}_{t}^{2}} \mathbf{y}_{\mathbf{M}^c,t}+\frac{\bar{\alpha}_{t-1} \beta_{t}^{2}}{\bar{\beta}_{t}^{2}} \mathbf{y}_{\mathbf{M}^c,0}, \tilde{\beta}_{t}=\frac{\beta_{t} \bar{\beta}_{t-1}}{\bar{\beta}_{t}}$. In MC-DDPM, we assume that $\alpha_{t}$ is set specially so that $\bar{\alpha}_{T} \approx 0$, i.e. $q\left( \mathbf{y}_{\mathbf{M}^c,T} \mid \mathbf{y}_{\mathbf{M}^c,0} \right) \approx \mathcal{N}\left(\mathbf{0}, \bar{\beta}_{T}^2\mathbf{M}^c \right)$ is a noise distribution independent of $\mathbf{y}_{M^c, 0}$.

Next, we discuss how to train MC-DDPM $p_{\theta}\left(\mathbf{y}_{\mathbf{M}^c,0}\mid \mathbf{M}^c, \mathbf{y}_{\mathbf{M}}\right)$. Firstly, let $p\left(\mathbf{y}_{\mathbf{M}^c, T}\mid  \mathbf{M}^c, \mathbf{y}_{\mathbf{M}}\right) = \mathcal{N}\left(\mathbf{0}, \bar{\beta}_{T}^2\mathbf{M}^c \right)$ so that it is nearly equal to $q\left( \mathbf{y}_{\mathbf{M}^c,T} \mid \mathbf{y}_{\mathbf{M}^c,0} \right)$. Training of $p_{\theta}\left(\mathbf{y}_{\mathbf{M}^c,0}\mid \mathbf{M}^c, \mathbf{y}_{\mathbf{M}}\right)$ is performed by optimizing the variational bound on negative log likelihood:
\begin{align*}
& \mathbb{E}\left[-\log p_{\theta}\left(\mathbf{y}_{\mathbf{M}^c,0}\mid \mathbf{M}^c, \mathbf{y}_{\mathbf{M}}\right)\right] \leq \mathbb{E}_{q}\left[-\log \frac{p_{\theta}\left(\mathbf{y}_{\mathbf{M}^c,0:T}\mid \mathbf{M}^c, \mathbf{y}_{\mathbf{M}}\right)}{q\left(\mathbf{y}_{\mathbf{M}^c,1:T} \mid \mathbf{y}_{\mathbf{M}^c,0}, \mathbf{M}^c, \mathbf{y}_{\mathbf{M}} \right)}\right] \notag \\
=& \mathbb{E}_{q}\left[-\log p\left(\mathbf{y}_{\mathbf{M}^c,T}\mid \mathbf{M}^c, \mathbf{y}_{\mathbf{M}}\right)-\sum_{t \geq 1} \log \frac{p_{\theta}\left(\mathbf{y}_{\mathbf{M}^c, t-1} \mid \mathbf{y}_{\mathbf{M}^c,t}, \mathbf{M}^c, \mathbf{y}_{\mathbf{M}}\right)}{q\left(\mathbf{y}_{\mathbf{M}^c, t} \mid \mathbf{y}_{\mathbf{M}^c,t-1}, \mathbf{M}^c, \mathbf{y}_{\mathbf{M}}\right)}\right]=: L. 
\end{align*}
Assuming that
\begin{equation}
    \boldsymbol{\mu}_{\theta}\left( \mathbf{y}_{\mathbf{M}^c,t}, t, \mathbf{M}^c, \mathbf{y}_{\mathbf{M}} \right)=\frac{1}{\alpha_{t}}\left(\mathbf{y}_{\mathbf{M}^c, t}-\frac{\beta_{t}^{2}}{\bar{\beta}_{t}} \boldsymbol{\varepsilon}_{\theta}\left(\mathbf{y}_{\mathbf{M}^c,t}, t, \mathbf{M}^c, \mathbf{y}_{\mathbf{M}}\right)\right),
\end{equation}
and supposing $\mathbf{y}_{\mathbf{M}^c, t} = \bar{\alpha}_{t} \mathbf{y}_{\mathbf{M}^c,0} + \boldsymbol{\varepsilon}, \boldsymbol{\varepsilon} \sim \mathcal{N} \left(\mathbf{0}, \bar{\beta}_{t}^{2} \mathbf{M}^c \right)$ (Eq.~\ref{eq:MC-DDPM_q_t}), after reweighting $L$ can be simplified as follows:
\begin{equation}
    L_{\mathrm{simple}}=\mathbb{E}_{\mathbf{y}^c_{\mathbf{M},0}, t, \boldsymbol{\varepsilon}} \left\| \boldsymbol{\varepsilon} -\boldsymbol{\varepsilon}_{\theta}\left(\bar{\alpha}_{t} \mathbf{y}^c_{\mathbf{M},0} + \bar{\beta}_{t} \boldsymbol{\varepsilon}, t, \mathbf{M}^c, \mathbf{y}_{\mathbf{M}}\right)\right\|_{2}^{2}, \boldsymbol{\varepsilon} \sim \mathcal{N} \left(\mathbf{0}, \mathbf{M}^c \right),
\end{equation}
where $t$ is uniform between $1$ and $T$. The details of derivation is in supplementary materials.
% By neglecting constants in $L$, we can derive the following training objective $L_{\mathrm{vlb}}$:
% \begin{equation}
% \label{eq:MC-DDPM_l_vlb}
%     L_{\mathrm{vlb}}=\mathbb{E}_{q, t}  \frac{1}{2 \sigma_{t}^{2}} \left\|\tilde{\boldsymbol{\mu}}_{t}-\boldsymbol{\mu}_{\theta}\left( \mathbf{y}_{\mathbf{M}^c,t}, t, \mathbf{M}^c, \mathbf{y}_{\mathbf{M}} \right)\right\|_{2}^{2}.
% \end{equation}
 
% , and combined with the formula of $\tilde{\boldsymbol{\mu}}$ in Eq.~\ref{eq:MC-DDPM_q_t-1_t_0}, $L_{\mathrm{vlb}}$ 
% \begin{equation*}
%     L_{\mathrm{vlb}}=\mathbb{E}_{\mathbf{y}^c_{\mathbf{M},0}, t, \boldsymbol{\varepsilon}}\frac{\beta_{t}^{4}}{2 \alpha_{t}^{2} \bar{\beta}_{t}^{2}\sigma_{t}^{2}}\left\|\boldsymbol{\varepsilon} -\boldsymbol{\varepsilon}_{\theta}\left(\bar{\alpha}_{t} \mathbf{y}^c_{\mathbf{M},0} + \bar{\beta}_{t} \boldsymbol{\varepsilon}, t, \mathbf{M}^c, \mathbf{y}_{\mathbf{M}}\right)\right\|_{2}^{2}, \boldsymbol{\varepsilon} \sim \mathcal{N} \left(\mathbf{0}, \mathbf{M}^c \right).
% \end{equation*}
% We also consider to train on the variant weighted objective as follows:

Algorithm~\ref{alg:MC-DDPM_training} displays the complete training procedure with this simplified objective and Algorithm~\ref{alg:MC-DDPM_sampling} shows the sampling process.

Since MC-DDPM can produce multiple samples of the posterior distribution $q\left(\mathbf{x} \mid \mathbf{y}_{\mathbf{M}}, \mathbf{M} \right)$, the pixel-variance can be computed by Monte Carlo approach which is used to quantify uncertainty of reconstruction. 

\begin{figure}[t]
\begin{minipage}[t]{0.495\textwidth}
\begin{algorithm}[H]
\caption{MC-DDPM Training}
\label{alg:MC-DDPM_training}
\begin{algorithmic}[1]
\Repeat
\State $\mathbf{x} \sim q \left( \mathbf{x} \right)$, obtain $\mathbf{M}$ and $\mathbf{M}^c$
\State $\mathbf{y}_{\mathbf{M}} = \mathbf{M} \mathbf{A} \mathbf{x} $, $\mathbf{y}_{\mathbf{M}^c} = \mathbf{M}^c \mathbf{A} \mathbf{x} $ 
\State $\boldsymbol{\varepsilon} \sim \mathcal{N} \left(\mathbf{0}, \mathbf{M}^c \right)$
\State $t \sim \operatorname{Uniform}(\{1, \ldots, T\})$
\State $\mathbf{y}^c_{\mathbf{M},t} = \bar{\alpha}_{t} \mathbf{y}^c_{\mathbf{M},0} + \bar{\beta}_{t} \boldsymbol{\varepsilon}$
\State Take gradient descent step on 
\Statex $\quad\quad\quad \nabla_{\theta} \left\| \boldsymbol{\varepsilon} -\boldsymbol{\varepsilon}_{\theta}\left(\mathbf{y}^c_{\mathbf{M},t}, t, \mathbf{M}^c, \mathbf{y}_{\mathbf{M}}\right)\right\|_{2}^{2}$
\Until{converged}
\end{algorithmic}
\end{algorithm} 
\end{minipage}
\hfill
\begin{minipage}[t]{0.495\textwidth}
\begin{algorithm}[H]
\caption{MC-DDPM Sampling}
\label{alg:MC-DDPM_sampling}
\begin{algorithmic}[1]
\State Given $\mathbf{M}^c$ and $\mathbf{y}_{\mathbf{M}}$
\State $\mathbf{y}_{\mathbf{M}^c, T} \sim \mathcal{N} \left(\mathbf{0}, \bar{\beta}_{T}^2 \mathbf{M}^c \right)$
\For{$t = T, ..., 1$}
\State $\mathbf{z}_{t} \sim \mathcal{N}(\mathbf{0}, \mathbf{M}^c)$ if $t > 1$, else $\mathbf{z}_{t} = \mathbf{0}$ 
\State $\boldsymbol{\mu}_{t} = \boldsymbol{\mu}_{\theta}\left( \mathbf{y}_{\mathbf{M}^c,t}, t, \mathbf{M}^c, \mathbf{y}_{\mathbf{M}} \right)$
\State $\mathbf{y}_{\mathbf{M}^c, t-1}= \boldsymbol{\mu}_{t} + \sigma_{t} \mathbf{z}_{t}$ 
\EndFor
\State \textbf{return} $\mathbf{x} = \mathbf{A}^{-1} \left(\mathbf{y}_{\mathbf{M}} + \mathbf{y}_{\mathbf{M}^c, 0}\right)$
\end{algorithmic}
\end{algorithm}
\end{minipage}
\end{figure}

\section{Experiments}
\label{sec:experiments}

We apply MC-DDPPM to accelerated MRI reconstruction where $\mathbf{A}$ is 2d Fourier transform and $\mathbf{y}_{\mathbf{M}}$ is the under-sampled k-space data. The specific design for $\boldsymbol{\varepsilon}_{\theta}\left(\mathbf{y}_{\mathbf{M}^c, t}, t, \mathbf{M}^c, \mathbf{y}_{\mathbf{M}}\right)$ in our experiments is given as follows:
\begin{equation}
\label{eq:MC-DDPM_mri_recon}
    \boldsymbol{\varepsilon}_{\theta}\left(\mathbf{y}_{\mathbf{M}^c, t}, t, \mathbf{M}^c, \mathbf{y}_{\mathbf{M}}\right) = \mathbf{M}^c f\left(g\left(\mathbf{A}^{-1}\left(\mathbf{y}_{\mathbf{M}^c, t} + \mathbf{y}_{M} \right), \mathbf{A}^{-1}\mathbf{y}_{M}\right), t; \theta \right),
\end{equation}
where $f$ is a deep neural network and $g\left(\cdot,\cdot \right)$ is the concatenation operation. Because MR image $\mathbf{x}$ is in complex filed, we use $\left|\mathbf{x}\right|$, the magnitude of it, as the final image. Pixel-wise variance is also computed using magnitude images.

\subsection{Experimental Setting}
\label{sec:experiments_setting}

All experiments are performed with fastMRI single-coil knee dataset \citep{zbontar2018fastmri}, which is publicly available\footnote{https://fastmri.org} and is divided into two parts, proton-density with (PDFS) and without fat suppression (PD). We trained the network with k-space data which were computed from $320 \times 320$ size complex images. We base the implementation of guided-DDPM \citep{dhariwal2021diffusion} and also follow similar setting for the diffusion process in \citep{dhariwal2021diffusion} but multiply $\beta_{t}$ by $0.5$ so that $\bar{\beta}_{T} \approx 0.5$. All networks were trained with learning rate of $0.0001$ using AdamW optimizer. More details of experiments is in supplementary materials.

To verify superiority, we perform comparison studies with baseline methods (U-Net \citep{ronneberger2015u}) used in \citep{zbontar2018fastmri}. The evaluation metrics, peak signal-to-noise ratio (PSNR) and structural similarity index (SSIM), of score-based reconstruction method proposed in \citep{chung2021score} are also used for comparison\footnote{They are provided in the paper of \citep{chung2021score}.} since their experiments are conducted on the same dataset.

\subsection{Experimental Results}
\label{sec:experiments_results}

We show the results of PD with $4 \times$ (the first row) and $8\times$ (the second row) acceleration in Fig.~\ref{fig:reconstruction}. More results are shown in supplementary materials. We compare our method to zero-filled reconstruction (ZF) and U-Net. Since MC-DDPM can produce multiple reconstruction samples, we use the mean of 20 samples as the object for comparison. We observe that the proposed method performs best both in $4 \times$ and $8 \times$ accelerations, where we see virtually more realistic structures and less error in the zoomed-in image than ZF and U-Net. In the last column of Fig.~\ref{fig:reconstruction}, we show the standard deviation of the samples. As the acceleration factor is increased, we see that the uncertainty increases correspondingly. Quantitative metrics in Table.~\ref{tab:evaluation_metrics} also confirm the superiority of our method. In the last two columns of Table.~\ref{tab:evaluation_metrics}, we compare MC-DDPM to the score-based reconstruction method proposed in \citep{chung2021score}. Because the testing volumes are randomly selected both in our experiments and in \citep{chung2021score}, it is impossible to compare directly. As a substitution, we compare the enhancement of evaluation metrics which is computed by the result of proposed method subtracting the result of U-Net. Due to the experiments in \citep{chung2021score} were conducted on the whole dataset (both PD and PDFS), we compute the average enhancement of PD and PDFS as our final 
result. Our method outperforms \citep{chung2021score} by 3.62/0.077 ($4 \times$) and 2.96/0.089 ($8 \times$) in PSNR/SSIM.

\begin{figure}[ht]
\begin{minipage}{0.45\linewidth}
    \centering
    \includegraphics[width=\linewidth]{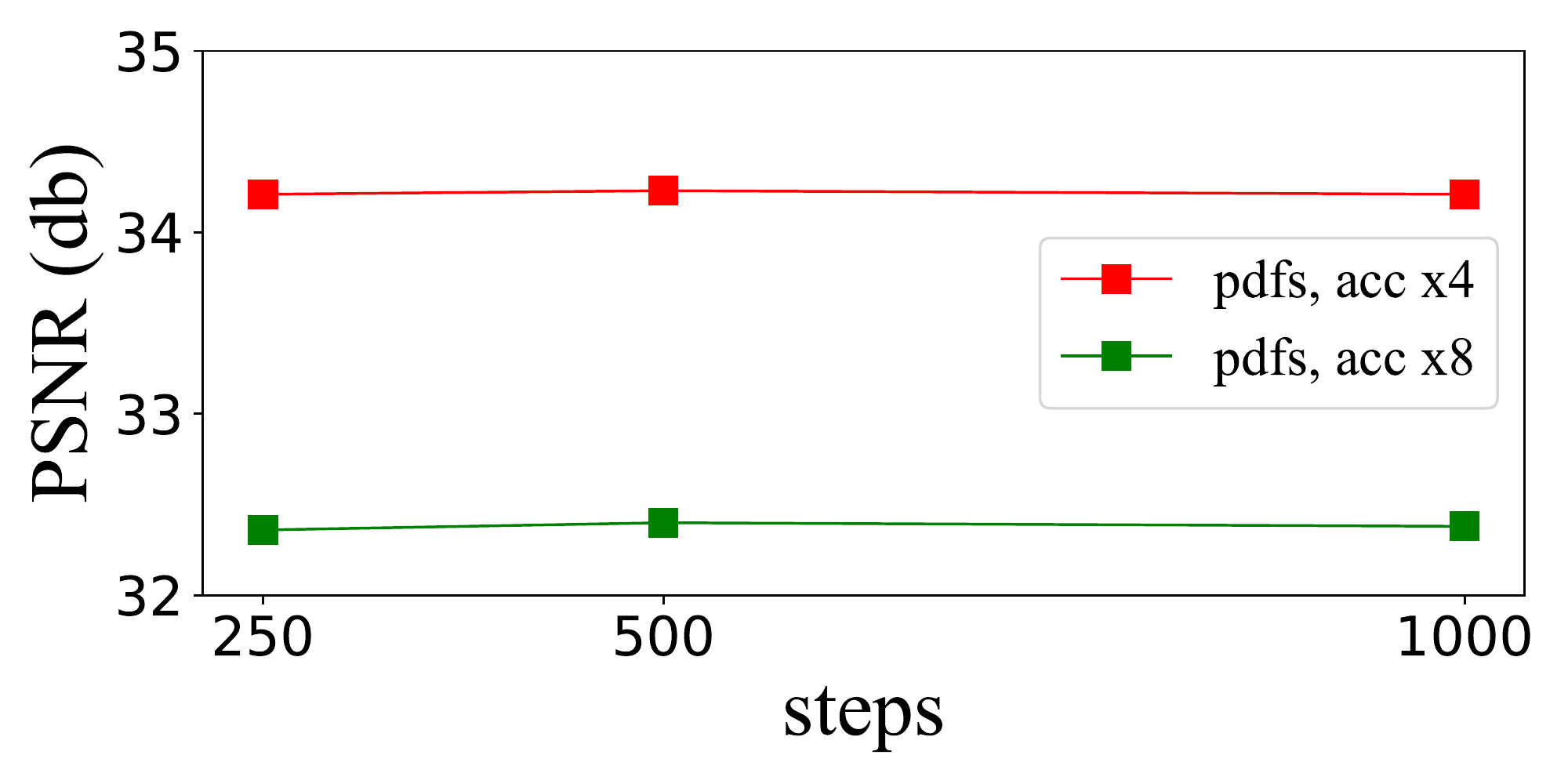}
    \caption{Tradeoff between number of sampling steps vs. PSNR testing on one volume of PDFS.}
    \label{fig:steps}
\end{minipage}
\hfill
\begin{minipage}{0.45\linewidth}
    \centering
    \includegraphics[width=\linewidth]{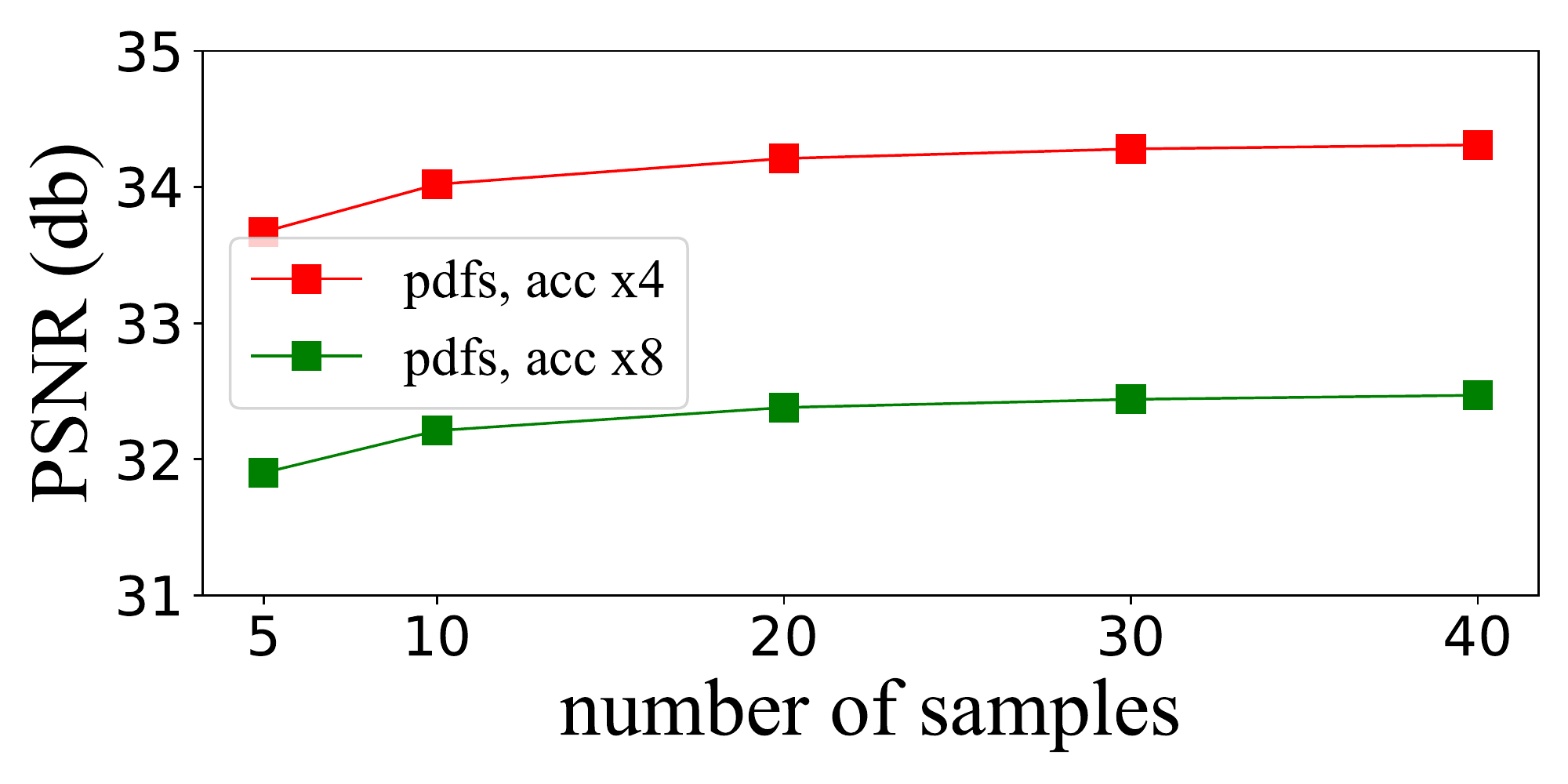}
    \caption{Tradeoff between number of samples vs. PSNR testing on one volume of PDFS.}
    \label{fig:num_samples}
\end{minipage}
\end{figure}

We also explore the effects of sampling steps and number of samples on reconstruction quality, which are illustrated in Fig.~\ref{fig:steps} and Fig.~\ref{fig:num_samples}. The two experiments are conducted on one volume of PDFS with $4 \times$ and $8 \times$ acceleration, and PSNR is computed on the mean of generated samples. We discover that: (1) even the sampling steps decrease to $250$, PSNR only reduces a little; (2) the quality of the mean of samples is enhanced when the number of samples increases and seems to converge. Taking the efficiency into account, 20 samples with 250 sampling steps may be a good choice.

\begin{figure}[t]
\centering
\includegraphics[width=\textwidth]{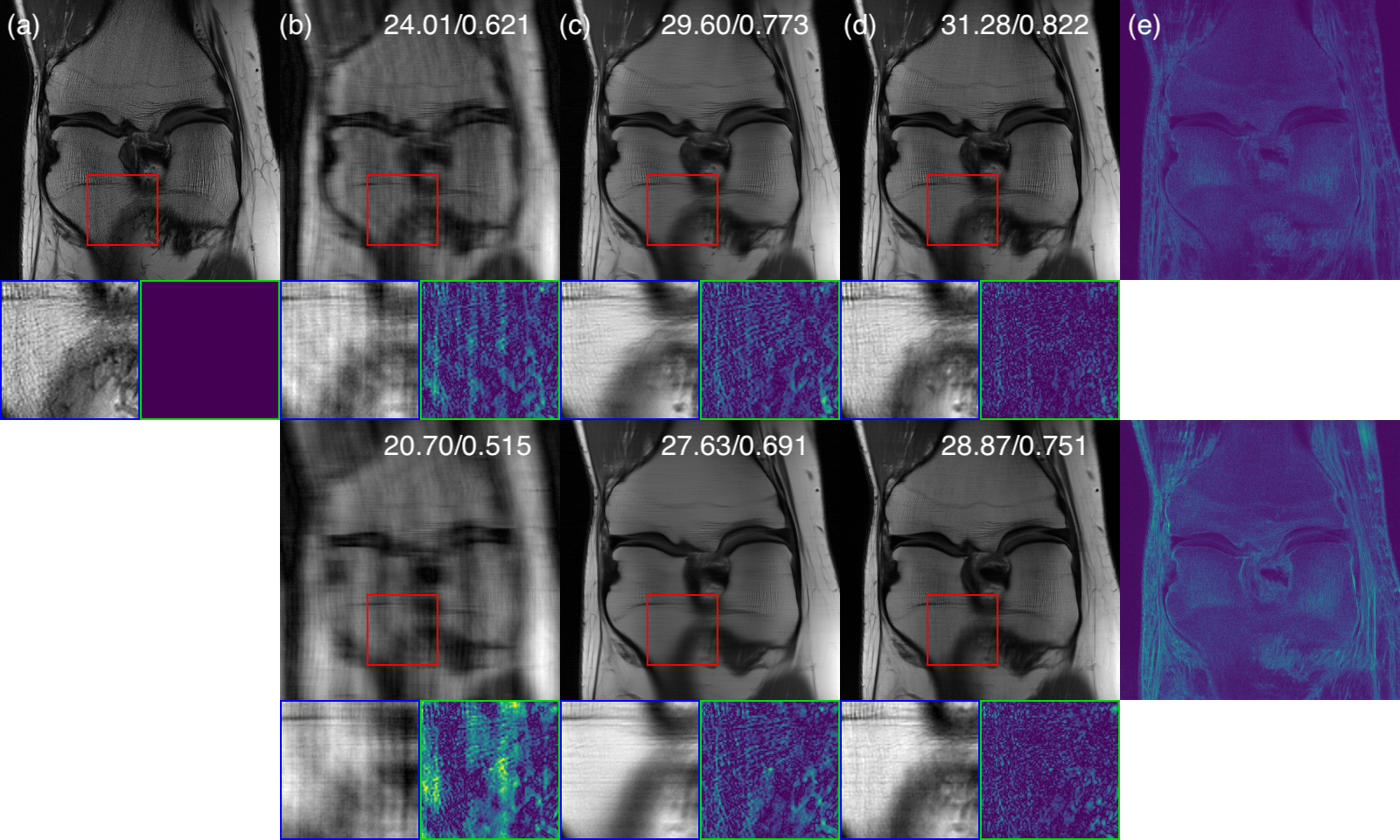}
\caption{Reconstruction results of $4 \times$ (the first row) and $8 \times$ (the second row) on PD data: (a) ground-truth, (b) zero-filled reconstruction, (c) U-Net, (d) the mean of samples generated by proposed method, (e) the standard deviation of the samples: range is set to [0, 0.2]. Blue box: Zoom in version of the indicated red box. Green box: Difference magnitude of the inset. White numbers indicate PSNR and SSIM, respectively.}
\label{fig:reconstruction}
\end{figure}

\begin{table}[t]
\caption{Quantitative metrics. Numbers in bold face indicate the best metric out of all the methods. The enhancement in the last two columns is computed based on U-Net.}
    \centering
    \begin{tabular}{c|c|p{0.9cm}<{\centering}p{0.9cm}<{\centering}p{0.9cm}<{\centering}|p{0.9cm}<{\centering}p{0.9cm}<{\centering}p{0.9cm}<{\centering}|p{1.7cm}<{\centering}p{0.9cm}<{\centering}}
    \hline
        ~ & ~ & \multicolumn{3}{c|}{PD} & \multicolumn{3}{c|}{PDFS} & \multicolumn{2}{c}{Enhancement} \\
        ~ & ~ & ZF & U-Net & Ours & ZF & U-Net & Ours & \cite{chung2021score} & Ours \\
        \hline
        \multirow{2}{*}{$\times$ 4} & PSNR & 29.62 & 34.04 & \textbf{36.69} & 26.32 & 28.30 & \textbf{33.00} & +0.06 & \textbf{+3.68}  \\
        ~ & SSIM & 0.745 & 0.834 & \textbf{0.905} & 0.545 & 0.648 & \textbf{0.735} & +0.002 & \textbf{+0.079}  \\
        \hline
        \multirow{2}{*}{$\times$ 8} & PSNR & 25.94 & 31.13 & \textbf{33.49} & 24.90 & 26.17 & \textbf{31.75} & +1.01 & \textbf{+3.97}  \\
        ~ & SSIM & 0.667 & 0.750 & \textbf{0.862} & 0.513 & 0.580 & \textbf{0.702} & +0.028 & \textbf{+0.117}  \\
    \hline
    \end{tabular}
    
    \label{tab:evaluation_metrics}
\end{table}

\subsection{Discussion}
\label{sec:experiments_discussion}

It is very common in medical imaging that the measurement is under sampled to reduce the cost or dosage. Therefore, it is important to define the conditional diffusion process in the measurement space for a reconstruction task. In this project, although our experiments are conducted using MR images, our method can be applied to other under-sampled medical image reconstruction tasks, such as limited angle or spare view CT Reconstruction. In our MC-DDPM, the variance schedule $\left\{\beta_{t}\right\}$ is an important hyper-parameter that is potentially related to the sampling quality and efficiency. Further investigation on hyper-parameter $\left\{\beta_{t}\right\}$ is planned in our future study.

% \section{Related Works}
% \label{sec:related_works}

\section{Conclusion}
\label{sec:conclusion}

In this paper we present a novel and unified mathematical framework, MC-DDPM, for medical image reconstruction using under-sampled measurements. Our method applies diffusion process in measurement domain with conditioned under-sampling mask, and provides an estimate of uncertainty as output. The superior performance of our method is demonstrated using accelerated MRI reconstruction, although MC-DDPM should potentially work for other under-sampled medical image reconstruction tasks.

\section{Acknowledgements} Thank Zifan Chen for his discussion and advice for paper writing.

% \bibliography{ref}

\begin{thebibliography}{26}
\providecommand{\natexlab}[1]{#1}
\providecommand{\url}[1]{\texttt{#1}}
\expandafter\ifx\csname urlstyle\endcsname\relax
  \providecommand{\doi}[1]{doi: #1}\else
  \providecommand{\doi}{doi: \begingroup \urlstyle{rm}\Url}\fi

\bibitem[Ho et~al.(2020)Ho, Jain, and Abbeel]{ho2020denoising}
Jonathan Ho, Ajay Jain, and Pieter Abbeel.
\newblock Denoising diffusion probabilistic models.
\newblock \emph{arXiv preprint arXiv:2006.11239}, 2020.

\bibitem[Sohl-Dickstein et~al.(2015)Sohl-Dickstein, Weiss, Maheswaranathan, and
  Ganguli]{sohl2015deep}
Jascha Sohl-Dickstein, Eric Weiss, Niru Maheswaranathan, and Surya Ganguli.
\newblock Deep unsupervised learning using nonequilibrium thermodynamics.
\newblock In \emph{International Conference on Machine Learning}, pages
  2256--2265. PMLR, 2015.

\bibitem[Song et~al.(2020)Song, Sohl-Dickstein, Kingma, Kumar, Ermon, and
  Poole]{song2020score}
Yang Song, Jascha Sohl-Dickstein, Diederik~P Kingma, Abhishek Kumar, Stefano
  Ermon, and Ben Poole.
\newblock Score-based generative modeling through stochastic differential
  equations.
\newblock \emph{arXiv preprint arXiv:2011.13456}, 2020.

\bibitem[Dhariwal and Nichol(2021)]{dhariwal2021diffusion}
Prafulla Dhariwal and Alex Nichol.
\newblock Diffusion models beat gans on image synthesis.
\newblock \emph{arXiv preprint arXiv:2105.05233}, 2021.

\bibitem[Xie et~al.(2022)Xie, Wu, Dong, and Li]{xie2022trained}
Yutong Xie, Dufan Wu, Bin Dong, and Quanzheng Li.
\newblock Trained model in supervised deep learning is a conditional risk
  minimizer.
\newblock \emph{arXiv preprint arXiv:2202.03674}, 2022.

\bibitem[Hyv{\"a}rinen and Dayan(2005)]{hyvarinen2005estimation}
Aapo Hyv{\"a}rinen and Peter Dayan.
\newblock Estimation of non-normalized statistical models by score matching.
\newblock \emph{Journal of Machine Learning Research}, 6\penalty0 (4), 2005.

\bibitem[Song and Ermon(2019)]{song2019generative}
Yang Song and Stefano Ermon.
\newblock Generative modeling by estimating gradients of the data distribution.
\newblock \emph{Advances in Neural Information Processing Systems}, 32, 2019.

\bibitem[Song et~al.(2021)Song, Shen, Xing, and Ermon]{song2021solving}
Yang Song, Liyue Shen, Lei Xing, and Stefano Ermon.
\newblock Solving inverse problems in medical imaging with score-based
  generative models.
\newblock \emph{arXiv preprint arXiv:2111.08005}, 2021.

\bibitem[Jalal et~al.(2021)Jalal, Arvinte, Daras, Price, Dimakis, and
  Tamir]{jalal2021robust}
Ajil Jalal, Marius Arvinte, Giannis Daras, Eric Price, Alex Dimakis, and
  Jonathan Tamir.
\newblock Robust compressed sensing mri with deep generative priors.
\newblock In \emph{Thirty-Fifth Conference on Neural Information Processing
  Systems}, 2021.

\bibitem[Chung et~al.(2021)]{chung2021score}
Hyungjin Chung et~al.
\newblock Score-based diffusion models for accelerated mri.
\newblock \emph{arXiv preprint arXiv:2110.05243}, 2021.

\bibitem[Zbontar et~al.(2018)Zbontar, Knoll, Sriram, Murrell, Huang, Muckley,
  Defazio, Stern, Johnson, Bruno, et~al.]{zbontar2018fastmri}
Jure Zbontar, Florian Knoll, Anuroop Sriram, Tullie Murrell, Zhengnan Huang,
  Matthew~J Muckley, Aaron Defazio, Ruben Stern, Patricia Johnson, Mary Bruno,
  et~al.
\newblock fastmri: An open dataset and benchmarks for accelerated mri.
\newblock \emph{arXiv preprint arXiv:1811.08839}, 2018.

\bibitem[Paszke et~al.(2019)Paszke, Gross, Massa, Lerer, Bradbury, Chanan,
  Killeen, Lin, Gimelshein, Antiga, et~al.]{paszke2019pytorch}
Adam Paszke, Sam Gross, Francisco Massa, Adam Lerer, James Bradbury, Gregory
  Chanan, Trevor Killeen, Zeming Lin, Natalia Gimelshein, Luca Antiga, et~al.
\newblock Pytorch: An imperative style, high-performance deep learning library.
\newblock \emph{Advances in neural information processing systems}, 32, 2019.

\bibitem[Ronneberger et~al.(2015)Ronneberger, Fischer, and
  Brox]{ronneberger2015u}
Olaf Ronneberger, Philipp Fischer, and Thomas Brox.
\newblock U-net: Convolutional networks for biomedical image segmentation.
\newblock In \emph{International Conference on Medical image computing and
  computer-assisted intervention}, pages 234--241. Springer, 2015.

\bibitem[Wang et~al.(2019)Wang, Zeng, Wang, and Guo]{wang2019admm}
Jiaxi Wang, Li~Zeng, Chengxiang Wang, and Yumeng Guo.
\newblock Admm-based deep reconstruction for limited-angle ct.
\newblock \emph{Physics in Medicine \& Biology}, 64\penalty0 (11):\penalty0
  115011, 2019.

\bibitem[Zhang et~al.(2019)Zhang, Dong, and Liu]{zhang2019jsr}
Haimiao Zhang, Bin Dong, and Baodong Liu.
\newblock Jsr-net: a deep network for joint spatial-radon domain ct
  reconstruction from incomplete data.
\newblock In \emph{ICASSP 2019-2019 IEEE International Conference on Acoustics,
  Speech and Signal Processing (ICASSP)}, pages 3657--3661. IEEE, 2019.

\bibitem[Han and Ye(2018)]{han2018framing}
Yoseob Han and Jong~Chul Ye.
\newblock Framing u-net via deep convolutional framelets: Application to
  sparse-view ct.
\newblock \emph{IEEE transactions on medical imaging}, 37\penalty0
  (6):\penalty0 1418--1429, 2018.

\bibitem[Han et~al.(2019)Han, Sunwoo, and Ye]{han2019k}
Yoseo Han, Leonard Sunwoo, and Jong~Chul Ye.
\newblock k-space deep learning for accelerated mri.
\newblock \emph{IEEE transactions on medical imaging}, 39\penalty0
  (2):\penalty0 377--386, 2019.

\bibitem[Aggarwal et~al.(2018)Aggarwal, Mani, and Jacob]{aggarwal2018modl}
Hemant~K Aggarwal, Merry~P Mani, and Mathews Jacob.
\newblock Modl: Model-based deep learning architecture for inverse problems.
\newblock \emph{IEEE transactions on medical imaging}, 38\penalty0
  (2):\penalty0 394--405, 2018.

\bibitem[Eo et~al.(2018)Eo, Jun, Kim, Jang, Lee, and Hwang]{eo2018kiki}
Taejoon Eo, Yohan Jun, Taeseong Kim, Jinseong Jang, Ho-Joon Lee, and Dosik
  Hwang.
\newblock Kiki-net: cross-domain convolutional neural networks for
  reconstructing undersampled magnetic resonance images.
\newblock \emph{Magnetic resonance in medicine}, 80\penalty0 (5):\penalty0
  2188--2201, 2018.

\bibitem[Hammernik et~al.(2018)Hammernik, Klatzer, Kobler, Recht, Sodickson,
  Pock, and Knoll]{hammernik2018learning}
Kerstin Hammernik, Teresa Klatzer, Erich Kobler, Michael~P Recht, Daniel~K
  Sodickson, Thomas Pock, and Florian Knoll.
\newblock Learning a variational network for reconstruction of accelerated mri
  data.
\newblock \emph{Magnetic resonance in medicine}, 79\penalty0 (6):\penalty0
  3055--3071, 2018.

\bibitem[Nichol and Dhariwal(2021)]{nichol2021improved}
Alexander~Quinn Nichol and Prafulla Dhariwal.
\newblock Improved denoising diffusion probabilistic models.
\newblock In \emph{International Conference on Machine Learning}, pages
  8162--8171. PMLR, 2021.

\bibitem[Kong et~al.(2020)Kong, Ping, Huang, Zhao, and
  Catanzaro]{kong2020diffwave}
Zhifeng Kong, Wei Ping, Jiaji Huang, Kexin Zhao, and Bryan Catanzaro.
\newblock Diffwave: A versatile diffusion model for audio synthesis.
\newblock \emph{arXiv preprint arXiv:2009.09761}, 2020.

\bibitem[Niu et~al.(2020)Niu, Song, Song, Zhao, Grover, and
  Ermon]{niu2020permutation}
Chenhao Niu, Yang Song, Jiaming Song, Shengjia Zhao, Aditya Grover, and Stefano
  Ermon.
\newblock Permutation invariant graph generation via score-based generative
  modeling.
\newblock In \emph{International Conference on Artificial Intelligence and
  Statistics}, pages 4474--4484. PMLR, 2020.

\bibitem[Choi et~al.(2021)Choi, Kim, Jeong, Gwon, and Yoon]{choi2021ilvr}
Jooyoung Choi, Sungwon Kim, Yonghyun Jeong, Youngjune Gwon, and Sungroh Yoon.
\newblock Ilvr: Conditioning method for denoising diffusion probabilistic
  models.
\newblock \emph{arXiv preprint arXiv:2108.02938}, 2021.

\bibitem[Saharia et~al.(2021)Saharia, Ho, Chan, Salimans, Fleet, and
  Norouzi]{saharia2021image}
Chitwan Saharia, Jonathan Ho, William Chan, Tim Salimans, David~J Fleet, and
  Mohammad Norouzi.
\newblock Image super-resolution via iterative refinement.
\newblock \emph{arXiv preprint arXiv:2104.07636}, 2021.

\bibitem[Meng et~al.(2021)Meng, Song, Song, Wu, Zhu, and Ermon]{meng2021sdedit}
Chenlin Meng, Yang Song, Jiaming Song, Jiajun Wu, Jun-Yan Zhu, and Stefano
  Ermon.
\newblock Sdedit: Image synthesis and editing with stochastic differential
  equations.
\newblock \emph{arXiv preprint arXiv:2108.01073}, 2021.

\end{thebibliography}
% \bibliographystyle{unsrtnat}

\section{Supplementary Materials}

\subsection{Derivation of Eq.~8, Eq.~9 and Training Objective in Sect.~3}

Before the derivation, we review the definition of $\bar{\alpha}_{t}$ and $\bar{\beta}_{t}$:
\begin{equation}
    \bar{\alpha}_{t} = \prod_{i=1}^{t} \alpha_{i}, \bar{\beta}_{t}^2 = \sum^{t}_{i=1} \frac{\bar{\alpha}_{t}^2}{\bar{\alpha}_{i}^2} \beta_{i}^2,
\end{equation}
where $\bar{\alpha}_{0} = 1$ and $\bar{\beta}_0 = 0$.

\paragraph{Eq.~8}
\begin{equation*}
    q\left(\mathbf{y}_{\mathbf{M}^c,t} \mid \mathbf{y}_{\mathbf{M}^c,0}, \mathbf{M}^c, \mathbf{y}_{\mathbf{M}} \right) = \mathcal{N}\left(\bar{\alpha}_{t} \mathbf{y}_{\mathbf{M}^c,0}, \bar{\beta}_{t}^{2} \mathbf{M}^c\right).
\end{equation*}

\begin{proof}
According to Eq.~7, we have that:
\begin{equation*}
    \mathbf{y}_{\mathbf{M}^c, t} = \alpha_t \mathbf{y}_{\mathbf{M}^c, t-1} + \beta_t \boldsymbol{\tau}_t, \boldsymbol{\tau}_t \sim \mathcal{N}\left(\mathbf{0}, \mathbf{M}^c  \right).
\end{equation*}
We can rewrite the equation above recursively:
\begin{align*}
    \mathbf{y}_{\mathbf{M}^c, t} & = \alpha_t \mathbf{y}_{\mathbf{M}^c, t-1} + \beta_t \boldsymbol{\tau}_t \\
    & = \alpha_t \left( \alpha_{t-1} \mathbf{y}_{\mathbf{M}^c, t-2} + \beta_{t-1} \boldsymbol{\tau}_{t-1} \right) + \beta_t \boldsymbol{\tau}_t \\
    & = \alpha_t \alpha_{t-1} \mathbf{y}_{\mathbf{M}^c, t-2} +  \alpha_t  \beta_{t-1} \boldsymbol{\tau}_{t-1}  + \beta_t \boldsymbol{\tau}_t \\
    & = \left( \prod_{i=t-1}^t \alpha_{i} \right) \mathbf{y}_{\mathbf{M}^c, t-2} + \sum^{t-1}_{i=t-1} \left(\prod_{j=i+1}^t \alpha_{j} \right) \beta_{i} \boldsymbol{\tau}_{i} + \beta_{t}\boldsymbol{\tau}_t \\
    & = \left( \prod_{i=1}^t \alpha_{i} \right) \mathbf{y}_{\mathbf{M}^c, 0} + \sum^{t-1}_{i=1} \left(\prod_{j=i+1}^t \alpha_{j} \right) \beta_{i} \boldsymbol{\tau}_{i} + \beta_{t}\boldsymbol{\tau}_t \\
    &= \bar{\alpha}_{t} \mathbf{y}_{\mathbf{M}^c, 0} + \sum^{t}_{i=1} \frac{\bar{\alpha}_{t}}{\bar{\alpha}_{i}}\beta_{i} \boldsymbol{\tau}_{i}
\end{align*}
Because $\boldsymbol{\tau}_{i}, i = 1, ..., t$ are all sampled from $\mathcal{N}\left(\mathbf{0}, \mathbf{M}^c \right)$ and independent of each other, $\sum^{t}_{i=1} \frac{\bar{\alpha}_{t}}{\bar{\alpha}_{i}}\beta_{i} \boldsymbol{\tau}_{i}$ can be regarded as sampled from $\mathcal{N}\left( \mathbf{0}, \bar{\beta}_{t}^2 \mathbf{M}^c \right)$ according to the definition of $\bar{\beta}_{t}$. Therefore, we have that:
\begin{equation*}
    \mathbf{y}_{\mathbf{M}^c, t} = \bar{\alpha}_{t} \mathbf{y}_{\mathbf{M}^c, 0} +  \bar{\beta}_{t}^2 \boldsymbol{\varepsilon}, \boldsymbol{\varepsilon} \sim \mathcal{N}\left(\mathbf{0}, \mathbf{M}^c \right),
\end{equation*}
i.e. 
\begin{equation*}
    q\left(\mathbf{y}_{\mathbf{M}^c,t} \mid \mathbf{y}_{\mathbf{M}^c,0}, \mathbf{M}^c, \mathbf{y}_{\mathbf{M}} \right) = \mathcal{N}\left(\bar{\alpha}_{t} \mathbf{y}_{\mathbf{M}^c,0}, \bar{\beta}_{t}^{2} \mathbf{M}^c\right).
\end{equation*}
\end{proof}

\paragraph{Eq.~9}
\begin{equation*}
    q\left(\mathbf{y}_{\mathbf{M}^c, t-1} \mid \mathbf{y}_{\mathbf{M}^c, t}, \mathbf{y}_{\mathbf{M}^c,0}, \mathbf{M}, \mathbf{y}_{\mathbf{M}}\right) = \mathcal{N}\left(\tilde{\boldsymbol{\mu}}_{t}, \tilde{\beta}_{t} \mathbf{M}^c\right),
\end{equation*}
where $\tilde{\boldsymbol{\mu}}_{t}=\frac{\alpha_{t} \bar{\beta}_{t-1}^{2}}{\bar{\beta}_{t}^{2}} \mathbf{y}_{\mathbf{M}^c,t}+\frac{\bar{\alpha}_{t-1} \beta_{t}^{2}}{\bar{\beta}_{t}^{2}} \mathbf{y}_{\mathbf{M}^c,0}, \tilde{\beta}_{t}=\frac{\beta_{t} \bar{\beta}_{t-1}}{\bar{\beta}_{t}}$.
\begin{proof}

Utilizing Bayesian Formula, we can derive that:
\begin{align*}
     & q\left(\mathbf{y}_{\mathbf{M}^c, t-1} \mid \mathbf{y}_{\mathbf{M}^c, t}, \mathbf{y}_{\mathbf{M}^c,0}, \mathbf{M}, \mathbf{y}_{\mathbf{M}}\right) \\
     = & \frac{q\left(\mathbf{y}_{\mathbf{M}^c, t-1}, \mathbf{y}_{\mathbf{M}^c, t}  \mid  \mathbf{y}_{\mathbf{M}^c,0}, \mathbf{M}, \mathbf{y}_{\mathbf{M}}\right) }{q\left( \mathbf{y}_{\mathbf{M}^c, t} \mid \mathbf{y}_{\mathbf{M}^c,0}, \mathbf{M}, \mathbf{y}_{\mathbf{M}}\right) }\\
     = &\frac{q\left( \mathbf{y}_{\mathbf{M}^c, t}  \mid \mathbf{y}_{\mathbf{M}^c, t-1}, \mathbf{y}_{\mathbf{M}^c,0}, \mathbf{M}, \mathbf{y}_{\mathbf{M}}\right) q\left(\mathbf{y}_{\mathbf{M}^c, t-1},  \mid  \mathbf{y}_{\mathbf{M}^c,0}, \mathbf{M}, \mathbf{y}_{\mathbf{M}}\right) }{q\left( \mathbf{y}_{\mathbf{M}^c, t} \mid \mathbf{y}_{\mathbf{M}^c,0}, \mathbf{M}, \mathbf{y}_{\mathbf{M}}\right) } \\
     = & \frac{q\left( \mathbf{y}_{\mathbf{M}^c, t}  \mid \mathbf{y}_{\mathbf{M}^c, t-1}, \mathbf{M}, \mathbf{y}_{\mathbf{M}}\right) q\left(\mathbf{y}_{\mathbf{M}^c, t-1},  \mid  \mathbf{y}_{\mathbf{M}^c,0}, \mathbf{M}, \mathbf{y}_{\mathbf{M}}\right) }{q\left( \mathbf{y}_{\mathbf{M}^c, t} \mid \mathbf{y}_{\mathbf{M}^c,0}, \mathbf{M}, \mathbf{y}_{\mathbf{M}}\right) }
\end{align*}
According to Eq.~7, Eq.~8, it is easy to write the explicit expressions for the three distributions in last equation. Since the components of $\mathbf{y}_{M^c, 0}, \mathbf{y}_{M^c, t-1}$ and $\mathbf{y}_{M^c, t}$ at under-sampled positions are all $0$, we only need to consider the non-sampled part of $q\left(\mathbf{y}_{\mathbf{M}^c, t-1} \mid \mathbf{y}_{\mathbf{M}^c, t}, \mathbf{y}_{\mathbf{M}^c,0}, \mathbf{M}, \mathbf{y}_{\mathbf{M}}\right)$. For convenience, we use $\mathbf{y}_i$ to represent the non-sampled part of $\mathbf{y}_{\mathbf{M}^c, i}$ and neglect the condition $\left( \mathbf{M}, \mathbf{y}_{\mathbf{M}} \right)$ in derivation. Then, we have:

\begin{equation*}
    q\left( \mathbf{y}_{\mathbf{M}^c, t}  \mid \mathbf{y}_{\mathbf{M}^c, t-1}, \mathbf{M}, \mathbf{y}_{\mathbf{M}}\right) = q \left(\mathbf{y}_{t} \mid \mathbf{y}_{t-1} \right) = \frac{1}{\sqrt{ \left( 2 \pi \right)^{d} \beta_{t}^{2d} }} \exp \left\{ - \frac{\left\| \mathbf{y}_{t} - \alpha_{t} \mathbf{y}_{t-1} \right\|^{2}_2 }{2 \beta_{t}^2} \right\};
\end{equation*}
\begin{equation*}
    q\left( \mathbf{y}_{\mathbf{M}^c, t-1}  \mid \mathbf{y}_{\mathbf{M}^c, 0}, \mathbf{M}, \mathbf{y}_{\mathbf{M}}\right) = q \left(\mathbf{y}_{t-1} \mid \mathbf{y}_{0} \right) = \frac{1}{\sqrt{ \left( 2 \pi \right)^{d} \bar{\beta}_{t-1}^{2d} }} \exp \left\{ - \frac{\left\| \mathbf{y}_{t-1} - \bar{\alpha}_{t-1} \mathbf{y}_{0} \right\|^{2}_2 }{2 \bar{\beta}_{t-1}^2} \right\};
\end{equation*}
\begin{equation*}
    q\left( \mathbf{y}_{\mathbf{M}^c, t}  \mid \mathbf{y}_{\mathbf{M}^c, 0}, \mathbf{M}, \mathbf{y}_{\mathbf{M}}\right) = q \left(\mathbf{y}_{t} \mid \mathbf{y}_{0} \right) = \frac{1}{\sqrt{ \left( 2 \pi \right)^{d} \bar{\beta}_{t}^{2d} }} \exp \left\{ - \frac{\left\| \mathbf{y}_{t} - \bar{\alpha}_{t} \mathbf{y}_{0} \right\|^{2}_2 }{2 \bar{\beta}_{t}^2} \right\};
\end{equation*}
where $d = n - m$. Next, we derive the distribution of $q\left( \mathbf{y}_{t-1} \mid  \mathbf{y}_{t},  \mathbf{y}_{0} \right)$ according to the three Gaussian distributions above.

Firstly, the coefficient is:
\begin{equation*}
    \frac{\frac{1}{\sqrt{ \left( 2 \pi \right)^{d} \beta_{t}^{2d} }}  \frac{1}{\sqrt{ \left( 2 \pi \right)^{d} \bar{\beta}_{t-1}^{2d} }} }{\frac{1}{\sqrt{ \left( 2 \pi \right)^{d} \bar{\beta}_{t}^{2d} }}} = \frac{1}{\sqrt{ \left( 2 \pi \right)^{d} \left(\frac{\beta_{t} \bar{\beta}_{t-1}}{\bar{\beta}_{t}}\right)^{2d}}}
\end{equation*}

Secondly, the exponential part is (neglecting the negative sign):
\begin{align*}
    & \frac{\left\| \mathbf{y}_{t} - \alpha_{t} \mathbf{y}_{t-1} \right\|^{2}_2 }{2 \beta_{t}^2} + \frac{\left\| \mathbf{y}_{t-1} - \bar{\alpha}_{t-1} \mathbf{y}_{0} \right\|^{2}_2 }{2 \bar{\beta}_{t-1}^2} - \frac{\left\| \mathbf{y}_{t} - \bar{\alpha}_{t} \mathbf{y}_{0} \right\|^{2}_2 }{2 \bar{\beta}_{t}^2} \\
    = & \frac{\bar{\beta}_{t-1}^{2} \bar{\beta}_{t}^{2} \left\|\mathbf{y}_{t} - \alpha_{t} \mathbf{y}_{t-1} \right\|^{2}_2 + \beta_{t}^{2} \bar{\beta}_{t}^{2} \left\|\mathbf{y}_{t-1} - \bar{\alpha}_{t-1} \mathbf{y}_{0} \right\|^{2}_2 - \beta_{t}^{2} \bar{\beta}_{t-1}^{2} \left\|\mathbf{y}_{t} - \bar{\alpha}_{t} \mathbf{y}_{0} \right\|^{2}_2}{2 \beta_{t}^{2} \bar{\beta}_{t-1}^{2} \bar{\beta}_{t}^{2}} \\
    = & \frac{a\left\|\mathbf{y}_{t-1}\right\|^{2}_2 + b \mathbf{y}_{t-1}^{T} \mathbf{y}_{t} + c \mathbf{y}_{t-1}^T \mathbf{y}_{0} + d \left\|\mathbf{y}_{t}\right\|^{2}_2 + e \mathbf{y}_{t}^T \mathbf{y}_{0} + f\left\|\mathbf{y}_{0}\right\|_2^{2}}{2 \beta_{t}^{2} \bar{\beta}_{t-1}^{2} \bar{\beta}_{t}^{2}},
\end{align*}
where
\begin{align*}
    & a = \bar{\beta}_{t-1}^{2} \bar{\beta}_{t}^{2} \alpha_{t}^{2} + \beta_{t, i}^{2} \bar{\beta}_{t}^{2}; \\
    & b = - 2 \bar{\beta}_{t-1}^{2} \bar{\beta}_{t}^{2} \alpha_{t}; \\
    & c= - 2 \beta_{t}^{2} \bar{\beta}_{t}^{2} \bar{\alpha}_{t-1}; \\
    & d = \bar{\beta}_{t-1}^{2} \bar{\beta}_{t}^{2} - \beta_{t}^{2} \bar{\beta}_{t-1}^{2}; \\
    & e = 2 \beta_{t}^{2} \bar{\beta}_{t-1}^{2} \bar{\alpha}_{t}; \\
    & f = \beta_{t}^{2} \bar{\beta}_{t}^{2} \bar{\alpha}_{t-1}^{2} - \beta_{t}^{2} \bar{\beta}_{t-1}^{2} \bar{\alpha}_{t}^{2}.
\end{align*}
According to the definitions of $\bar{\alpha}_{t}$ and $\bar{\beta}_{t}$, we can derive the following useful equation:
\begin{equation*}
    \bar{\beta}_{t-1}^{2} \alpha_{t}^{2} + \beta_{t}^{2} = \bar{\beta}_{t}^{2}.
\end{equation*}
The equation above can be used to simplify $a, d$ and $f$. We have the following results:
\begin{equation*}
    a = \bar{\beta}_{t}^{4}, d = \alpha_{t}^{2} \bar{\beta}_{t-1}^{4}, f = \bar{\alpha}_{t-1}^{2} \beta_{t}^{4}.
\end{equation*}
Therefore, we derive that:
\begin{align*}
    & \frac{a\left\|\mathbf{y}_{t-1}\right\|^{2}_2 + b \mathbf{y}_{t-1}^{T} \mathbf{y}_{t} + c \mathbf{y}_{t-1}^T \mathbf{y}_{0} + d \left\|\mathbf{y}_{t}\right\|^{2}_2 + e \mathbf{y}_{t}^T \mathbf{y}_{0} + f\left\|\mathbf{y}_{0}\right\|_2^{2}}{2 \beta_{t}^{2} \bar{\beta}_{t-1}^{2} \bar{\beta}_{t}^{2}} \\
    = & \frac{\bar{\beta}_{t}^{4} \left\|\mathbf{y}_{t-1}\right\|^{2}_2 - 2 \bar{\beta}_{t-1}^{2} \bar{\beta}_{t}^{2} \alpha_{t} \mathbf{y}_{t-1}^{T} \mathbf{y}_{t}  - 2 \beta_{t}^{2} \bar{\beta}_{t}^{2} \bar{\alpha}_{t-1} \mathbf{y}_{t-1}^T \mathbf{y}_{0} + \alpha_{t}^{2} \bar{\beta}_{t-1}^{4} \left\|\mathbf{y}_{t}\right\|^{2}_2 + 2 \beta_{t}^{2} \bar{\beta}_{t-1}^{2} \bar{\alpha}_{t} \mathbf{y}_{t}^T \mathbf{y}_{0} + \bar{\alpha}_{t-1}^{2} \beta_{t}^{4} \left\|\mathbf{y}_{0}\right\|_2^{2}}{2 \beta_{t}^{2} \bar{\beta}_{t-1}^{2} \bar{\beta}_{t}^{2}} \\
    = & \frac{\bar{\beta}_{t}^{4} \left\|\mathbf{y}_{t-1}\right\|^{2}_2 - 2 \bar{\beta}_{t}^{2} \mathbf{y}_{t-1}^T \left( \alpha_{t} \bar{\beta}_{t-1}^{2} \mathbf{y}_{t} + \bar{\alpha}_{t-1} \beta_{t}^{2} \mathbf{y}_{0} \right) + \left\| \alpha_{t} \bar{\beta}_{t-1}^{2} \mathbf{y}_{t} + \bar{\alpha}_{t-1} \beta_{t}^{2} \mathbf{y}_{0} \right\|_2^{2}}{2 \beta_{t}^{2} \bar{\beta}_{t-1}^{2} \bar{\beta}_{t}^{2}} \\
    = & \frac{\left\|\bar{\beta}_{t}^{2} \mathbf{y}_{t-1} - \left(\alpha_{t} \bar{\beta}_{t-1}^{2} \mathbf{y}_{t} + \bar{\alpha}_{t-1} \beta_{t}^{2} \mathbf{y}_{0}\right) \right\|_2^2}{2 \beta_{t}^{2} \bar{\beta}_{t-1}^{2} \bar{\beta}_{t}^{2}} \\
    = &  \frac{\left\|\mathbf{y}_{t-1}- \left( \frac{\alpha_{t} \bar{\beta}_{t-1}^{2}}{\bar{\beta}_{t}^{2}} \mathbf{y}_{t} + \frac{\bar{\alpha}_{t-1} \beta_{t}^{2}}{\bar{\beta}_{t}^{2}} \mathbf{y}_{0} \right) \right\|_2^{2}}{2\frac{\beta_{t}^{2} \bar{\beta}_{t-1}^{2}}{\bar{\beta}_{t}^{2}}} .
\end{align*}
Finally, we obtain that:
\begin{align*}
    & q\left(\mathbf{y}_{\mathbf{M}^c, t-1} \mid \mathbf{y}_{\mathbf{M}^c, t}, \mathbf{y}_{\mathbf{M}^c,0}, \mathbf{M}, \mathbf{y}_{\mathbf{M}}\right) = q\left( \mathbf{y}_{t-1} \mid \mathbf{y}_{t}, \mathbf{y}_{0} \right) \\
    = & \frac{1}{\sqrt{ \left( 2 \pi \right)^{d} \left(\frac{\beta_{t} \bar{\beta}_{t-1}}{\bar{\beta}_{t}}\right)^{2d}}} \exp \left\{- \frac{\left\|\mathbf{y}_{t-1}- \left( \frac{\alpha_{t} \bar{\beta}_{t-1}^{2}}{\bar{\beta}_{t}^{2}} \mathbf{y}_{t} + \frac{\bar{\alpha}_{t-1} \beta_{t}^{2}}{\bar{\beta}_{t}^{2}} \mathbf{y}_{0} \right) \right\|_2^{2}}{2\frac{\beta_{t}^{2} \bar{\beta}_{t-1}^{2}}{\bar{\beta}_{t}^{2}}} \right\}.
\end{align*}
i.e.
\begin{equation*}
    q\left(\mathbf{y}_{\mathbf{M}^c, t-1} \mid \mathbf{y}_{\mathbf{M}^c, t}, \mathbf{y}_{\mathbf{M}^c,0}, \mathbf{M}, \mathbf{y}_{\mathbf{M}}\right) = \mathcal{N}\left(\tilde{\boldsymbol{\mu}}_{t}, \tilde{\beta}_{t} \mathbf{M}^c\right),
\end{equation*}
where $\tilde{\boldsymbol{\mu}}_{t}=\frac{\alpha_{t} \bar{\beta}_{t-1}^{2}}{\bar{\beta}_{t}^{2}} \mathbf{y}_{\mathbf{M}^c,t}+\frac{\bar{\alpha}_{t-1} \beta_{t}^{2}}{\bar{\beta}_{t}^{2}} \mathbf{y}_{\mathbf{M}^c,0}, \tilde{\beta}_{t}=\frac{\beta_{t} \bar{\beta}_{t-1}}{\bar{\beta}_{t}}$.
\end{proof}

\paragraph{Training Objective}

In DDPM \citep{ho2020denoising}, the training of $p_{\theta}\left(\mathbf{x}_{0} \right)$ is performed by optimizing the variational bound on negative log likelihood:
\begin{align*}
\mathbb{E}\left[-\log p_{\theta}\left(\mathbf{x}_{0}\right)\right] &\leq \mathbb{E}_{q}\left[-\log \frac{p_{\theta}\left(\mathbf{x}_{0: T}\right)}{q\left(\mathbf{x}_{1: T} \mid \mathbf{x}_{0}\right)}\right] \notag \\
&=\mathbb{E}_{q}\left[-\log p\left(\mathbf{x}_{T}\right)-\sum_{t \geq 1} \log \frac{p_{\theta}\left(\mathbf{x}_{t-1} \mid \mathbf{x}_{t}\right)}{q\left(\mathbf{x}_{t} \mid \mathbf{x}_{t-1}\right)}\right]=: L. 
\end{align*}
Similarly we can define $L$ for MC-DDPM as follows:
\begin{align*}
& \mathbb{E}\left[-\log p_{\theta}\left(\mathbf{y}_{\mathbf{M}^c,0}\mid \mathbf{M}^c, \mathbf{y}_{\mathbf{M}}\right)\right] \leq \mathbb{E}_{q}\left[-\log \frac{p_{\theta}\left(\mathbf{y}_{\mathbf{M}^c,0:T}\mid \mathbf{M}^c, \mathbf{y}_{\mathbf{M}}\right)}{q\left(\mathbf{y}_{\mathbf{M}^c,1:T} \mid \mathbf{y}_{\mathbf{M}^c,0}, \mathbf{M}^c, \mathbf{y}_{\mathbf{M}} \right)}\right] \notag \\
=& \mathbb{E}_{q}\left[-\log p\left(\mathbf{y}_{\mathbf{M}^c,T}\mid \mathbf{M}^c, \mathbf{y}_{\mathbf{M}}\right)-\sum_{t \geq 1} \log \frac{p_{\theta}\left(\mathbf{y}_{\mathbf{M}^c, t-1} \mid \mathbf{y}_{\mathbf{M}^c,t}, \mathbf{M}^c, \mathbf{y}_{\mathbf{M}}\right)}{q\left(\mathbf{y}_{\mathbf{M}^c, t} \mid \mathbf{y}_{\mathbf{M}^c,t-1}, \mathbf{M}^c, \mathbf{y}_{\mathbf{M}}\right)}\right]=: L. 
\end{align*}
Removing the constants in $L$, we can derive $L_{\mathrm{vlb}}$ as follows:
\begin{equation*}
    L_{\mathrm{vlb}}=\mathbb{E}_{q, t}  \frac{1}{2 \sigma_{t}^{2}} \left\|\tilde{\boldsymbol{\mu}}_{t}-\boldsymbol{\mu}_{\theta}\left( \mathbf{y}_{\mathbf{M}^c,t}, t, \mathbf{M}^c, \mathbf{y}_{\mathbf{M}} \right)\right\|_{2}^{2},
\end{equation*}
where $t$ is uniform between $1$ and $T$. Assuming that
\begin{equation}
\label{eq:MC-DDPM_l_vlb2}
    \boldsymbol{\mu}_{\theta}\left( \mathbf{y}_{\mathbf{M}^c,t}, t, \mathbf{M}^c, \mathbf{y}_{\mathbf{M}} \right)=\frac{1}{\alpha_{t}}\left(\mathbf{y}_{\mathbf{M}^c, t}-\frac{\beta_{t}^{2}}{\bar{\beta}_{t}} \boldsymbol{\varepsilon}_{\theta}\left(\mathbf{y}_{\mathbf{M}^c,t}, t, \mathbf{M}^c, \mathbf{y}_{\mathbf{M}}\right)\right),
\end{equation}
and supposing $\mathbf{y}_{\mathbf{M}^c, t} = \bar{\alpha}_{t} \mathbf{y}_{\mathbf{M}^c,0} + \boldsymbol{\varepsilon}, \boldsymbol{\varepsilon} \sim \mathcal{N} \left(\mathbf{0}, \bar{\beta}_{t}^{2} \mathbf{M}^c \right)$ (Eq.~8), $L_{\mathrm{vlb}}$ can be simplified as follows:
\begin{equation*}
    L_{\mathrm{vlb}}=\mathbb{E}_{\mathbf{y}^c_{\mathbf{M},0}, t, \boldsymbol{\varepsilon}}\frac{\beta_{t}^{4}}{2 \alpha_{t}^{2} \bar{\beta}_{t}^{2}\sigma_{t}^{2}}\left\|\boldsymbol{\varepsilon} -\boldsymbol{\varepsilon}_{\theta}\left(\bar{\alpha}_{t} \mathbf{y}^c_{\mathbf{M},0} + \bar{\beta}_{t} \boldsymbol{\varepsilon}, t, \mathbf{M}^c, \mathbf{y}_{\mathbf{M}}\right)\right\|_{2}^{2}, \quad \boldsymbol{\varepsilon} \sim \mathcal{N} \left(\mathbf{0}, \mathbf{M}^c \right).
\end{equation*}
After reweighting $L_{\mathrm{vlb}}$ we obtain the final training objective as follows:
\begin{equation*}
    L_{\mathrm{simple}}=\mathbb{E}_{\mathbf{y}^c_{\mathbf{M},0}, t, \boldsymbol{\varepsilon}} \left\| \boldsymbol{\varepsilon} -\boldsymbol{\varepsilon}_{\theta}\left(\bar{\alpha}_{t} \mathbf{y}^c_{\mathbf{M},0} + \bar{\beta}_{t} \boldsymbol{\varepsilon}, t, \mathbf{M}^c, \mathbf{y}_{\mathbf{M}}\right)\right\|_{2}^{2}, \boldsymbol{\varepsilon} \sim \mathcal{N} \left(\mathbf{0}, \mathbf{M}^c \right).
\end{equation*}

\subsection{More Details of Experiments}

For all volumes of training data, we drop the first and last five slices to avoid training the model with noise-only data as \citep{chung2021score} did. For testing, we randomly select 6 volumes from the validation set and dropped the first and last 5 slices from each volume for both PD and PDFS. The model architectures used in experiments stems from U-Net \citep{ronneberger2015u} and is added by time embedding modules and self-attention layers. We train the model with batch size of 48 for 35k steps in MC-DDPM experiments. About $\alpha_t$ and $\beta_t$, we first use the "cosine" schedule which is also used in \citep{dhariwal2021diffusion}, and then multiply $\beta_t$ by $0.5$. The details can be seen in our code. All code was implemented in PyTorch \citep{paszke2019pytorch}.

\subsection{More Experimental Results}

\begin{figure}
    \centering
    \includegraphics[width=\linewidth]{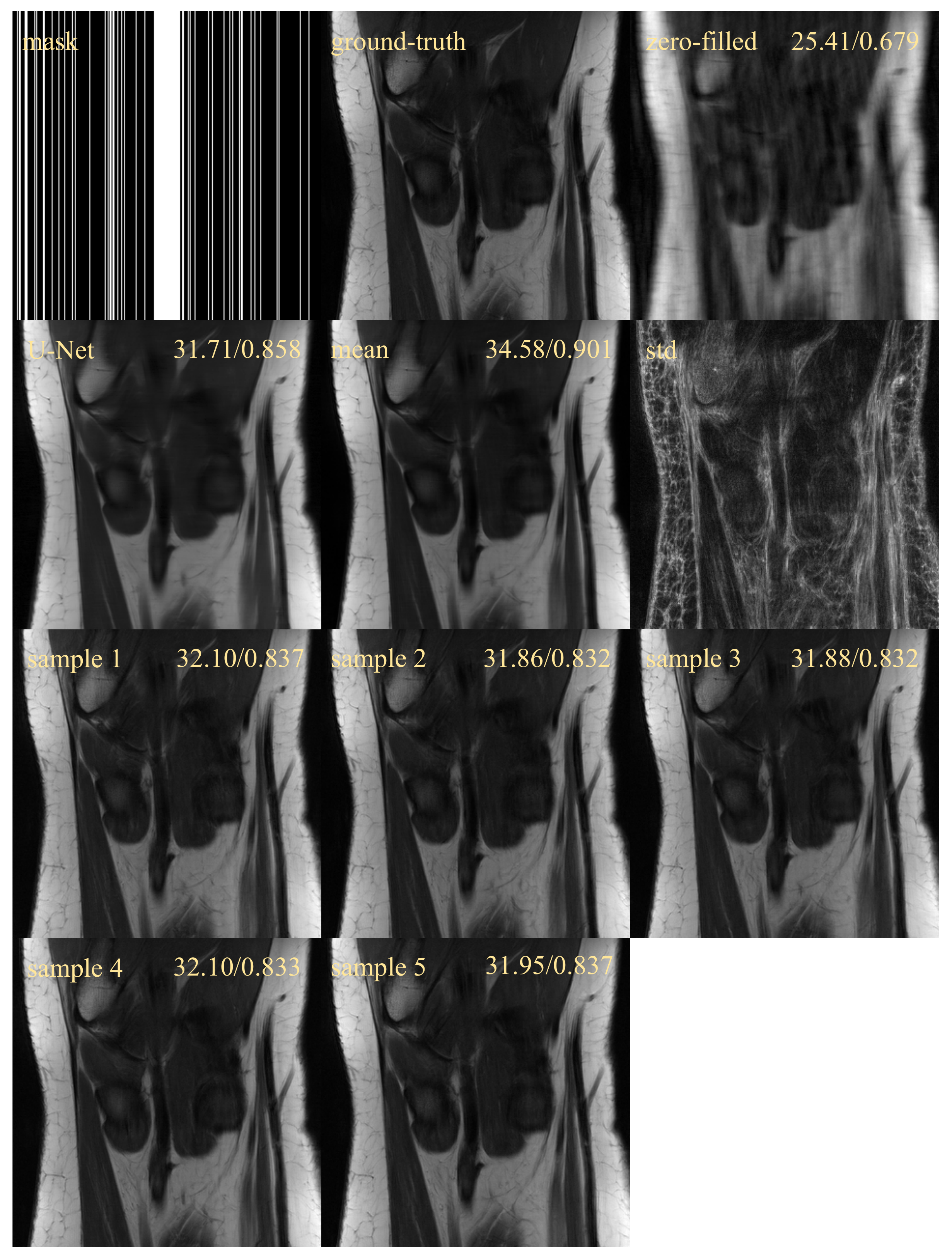}
    \caption{Reconstruction results of $4 \times$ on PD. The "mean" and "std" (ragne is set [0, 1]) are computed by 20 samples generated by proposed method. We also list five samples. Yellow numbers indicate PSNR and SSIM, respectively.}
    \label{fig:pd4x_samples}
\end{figure}

\begin{figure}
    \centering
    \includegraphics[width=\linewidth]{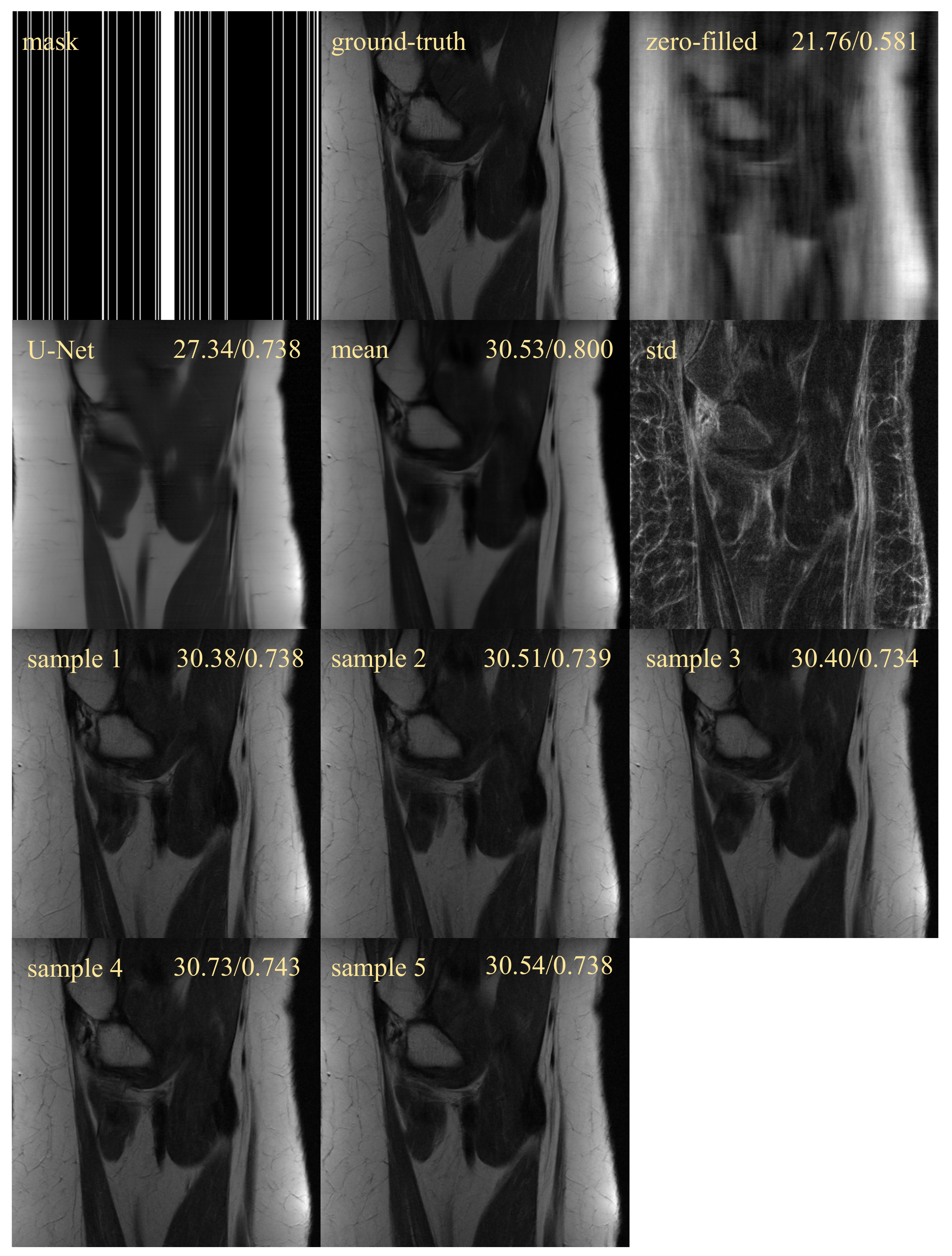}
    \caption{Reconstruction results of $8 \times$ on PD. The "mean" and "std" (ragne is set [0, 1]) are computed by 20 samples generated by proposed method. We also list five samples. Yellow numbers indicate PSNR and SSIM, respectively.}
    \label{fig:pd8x_samples}
\end{figure}

\begin{figure}
    \centering
    \includegraphics[width=\linewidth]{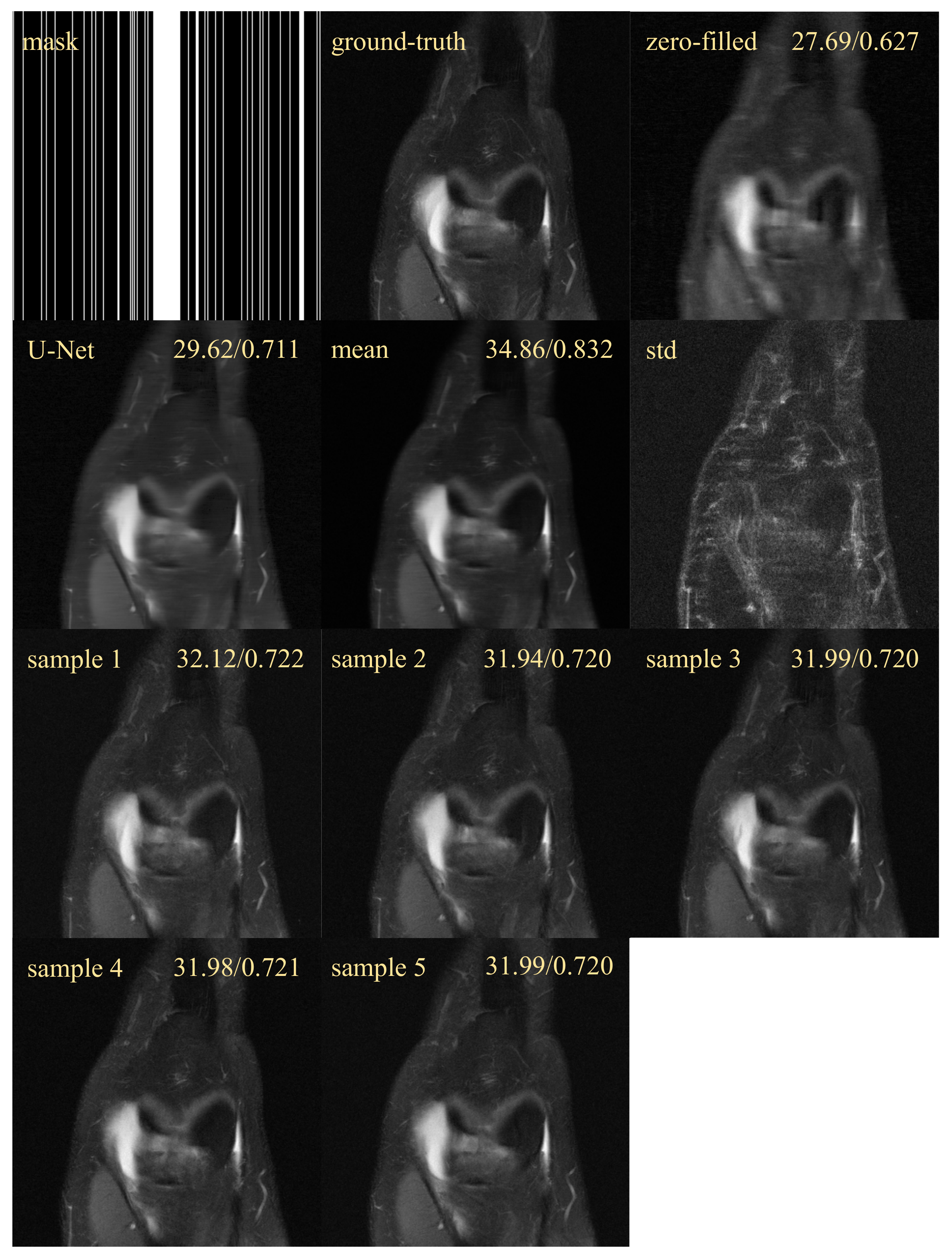}
    \caption{Reconstruction results of $4 \times$ on PDFS. The "mean" and "std" (ragne is set [0, 1]) are computed by 20 samples generated by proposed method. We also list five samples. Yellow numbers indicate PSNR and SSIM, respectively.}
    \label{fig:pdfs4x_samples}
\end{figure}

\begin{figure}
    \centering
    \includegraphics[width=\linewidth]{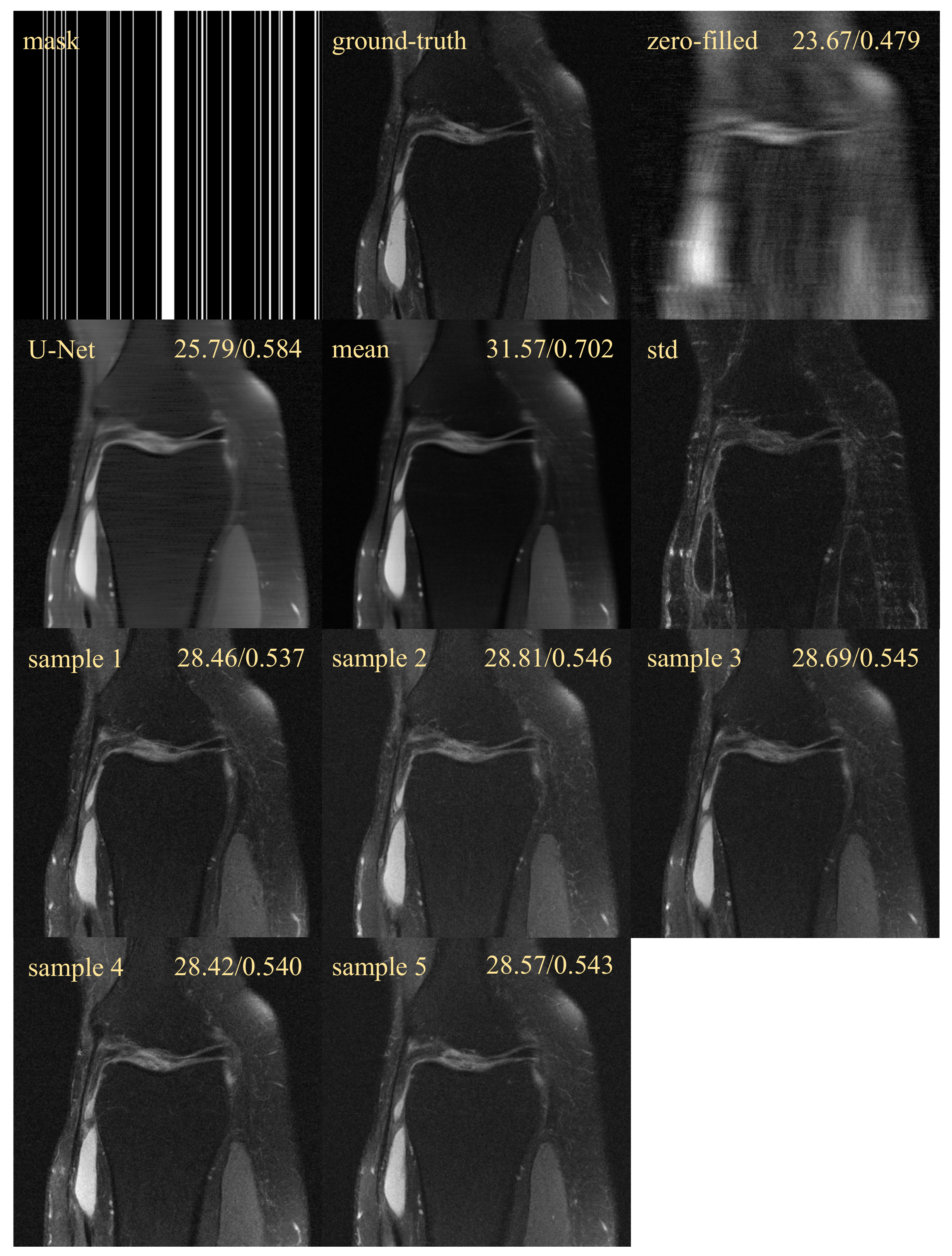}
    \caption{Reconstruction results of $8 \times$ on PDFS. The "mean" and "std" (ragne is set [0, 1]) are computed by 20 samples generated by proposed method. We also list five samples. Yellow numbers indicate PSNR and SSIM, respectively.}
    \label{fig:pdfs8x_samples}
\end{figure}

We list more quantitative metrics, including NMSE and MSE, for both volumes and slices in Table.~\ref{tab:more_evaluation_metrics_volume} and Table.~\ref{tab:more_evaluation_metrics_slice}. More reconstruction results are shown in Fig.~\ref{fig:pd4x_samples}, Fig.~\ref{fig:pd8x_samples}, Fig.~\ref{fig:pdfs4x_samples}, and Fig.~\ref{fig:pdfs8x_samples}.

\begin{table}[t]
\caption{Quantitative metrics (PSNR, SSIM, NMSE, MSE) which are computed on volumes. Numbers in bold face indicate the best metric out of all the methods.}
    \centering
    \begin{tabular}{c|c|ccc|ccc}
    \hline
        ~ & ~ & \multicolumn{3}{c|}{PD} & \multicolumn{3}{c}{PDFS} \\
        ~ & ~ & ZF & U-Net & Ours & ZF & U-Net & Ours \\
        \hline
        \multirow{4}{*}{$\times$ 4} & PSNR & 29.62 & 34.04 & \textbf{36.69} & 26.32 & 28.30 & \textbf{33.00}  \\
        ~ & SSIM & 0.745 & 0.834 & \textbf{0.905} & 0.545 & 0.648 & \textbf{0.735} \\
        ~ & NMSE & 0.0271 & 0.0107 & \textbf{0.0057} & 0.0776 & 0.0510 & \textbf{0.0164} \\
        ~ & MSE & 2.31e-10 & 7.64e-11 & \textbf{4.26e-11} & 5.11e-11 & 3.38e-11 & \textbf{1.08e-11} \\
        \hline
        \multirow{4}{*}{$\times$ 8} & PSNR & 25.94 & 31.13 & \textbf{33.49} & 24.90 & 26.17 & \textbf{31.75}  \\
        ~ & SSIM & 0.667 & 0.750 & \textbf{0.862} & 0.513 & 0.580 & \textbf{0.702} \\
        ~ & NMSE & 0.0630 & 0.0203 & \textbf{0.0114} & 0.105 & 0.0814 & \textbf{0.0215} \\
        ~ & MSE & 5.66e-10 & 1.50e-10 & \textbf{9.00e-11} & 6.88e-11 & 6.88e-11 & \textbf{1.40e-11} \\
    \hline
    \end{tabular}
    
    \label{tab:more_evaluation_metrics_volume}
\end{table}

\begin{table}[t]
\caption{Quantitative metrics (PSNR, SSIM, NMSE, MSE) which are computed on slices. Numbers in bold face indicate the best metric out of all the methods.}
    \centering
    \begin{tabular}{c|c|ccc|ccc}
    \hline
        ~ & ~ & \multicolumn{3}{c|}{PD} & \multicolumn{3}{c}{PDFS} \\
        ~ & ~ & ZF & U-Net & Ours & ZF & U-Net & Ours \\
        \hline
        \multirow{4}{*}{$\times$ 4} & PSNR & 27.15 & 31.32 & \textbf{33.98} & 23.07 & 25.15 & \textbf{29.76}  \\
        ~ & SSIM & 0.678 & 0.785 & \textbf{0.867} & 0.438 & 0.536 & \textbf{0.614} \\
        ~ & NMSE & 0.0272 & 0.0118 & \textbf{0.0060} & 0.0898 & 0.0611 & \textbf{0.0174} \\
        ~ & MSE & 2.33e-10 & 7.71e-11 & \textbf{4.32e-11} & 5.29e-11 & 3.53e-11 & \textbf{1.11e-11} \\
        \hline
        \multirow{4}{*}{$\times$ 8} & PSNR & 23.51 & 28.46 & \textbf{30.97} & 21.70 & 23.01 & \textbf{28.60}  \\
        ~ & SSIM & 0.592 & 0.695 & \textbf{0.813} & 0.399 & 0.468 & \textbf{0.571} \\
        ~ & NMSE & 0.0624 & 0.0222 & \textbf{0.0118} & 0.120 & 0.0961 & \textbf{0.0222} \\
        ~ & MSE & 5.29e-11 & 3.53e-11 & \textbf{1.11e-11} & 7.09e-11 & 5.58e-11 & \textbf{1.43e-11} \\
    \hline
    \end{tabular}
    
    \label{tab:more_evaluation_metrics_slice}
\end{table}

\newpage

\end{document}